\documentclass[10pt, conference, letterpaper]{IEEEtran}

\usepackage{epsfig}
\usepackage{amssymb}
\usepackage{amsmath}
\usepackage{amsfonts}
\usepackage{amsmath}
\usepackage{cases}
\usepackage{graphicx}
\usepackage{makeidx} 

\usepackage{url}
\usepackage{cite}
\usepackage{times}
\usepackage{graphicx}
\usepackage{graphics}
\usepackage{subfigure}
\usepackage{latexsym}
\usepackage{paralist}
\usepackage{xspace}
\usepackage{mathrsfs}
\usepackage{color}
\usepackage[ruled]{algorithm}
\usepackage[noend]{algorithmic}
\usepackage{mathtools}

\usepackage{url,epsfig,array}
\usepackage{tikz}
\usepackage{caption}

\allowdisplaybreaks 
\def\ie{\textit{i.e.}\xspace}

\def\etc{\textit{etc.}}

\def\eg{\textit{e.g.}\xspace}

\def\ourprotocol{\textit{SouTu} \xspace}
\def\ourprotocolNSP{\textit{SouTu}}

\newcommand{\CUTXY}[1]{{}}
\newcommand{\CUTTaeho}[1]{{}}

\newcommand{\code}{\texttt{GI}}

\newcommand{\en}{\texttt{AES.E}}

\newcommand{\garbled}{\texttt{G}}
\newcommand{\abe}{\texttt{ABE.E}}
\newcommand{\dabe}{\texttt{ABE.D}}
\newcommand{\enh}{\texttt{HE.E}}
\newcommand{\kg}{\texttt{KG}}
\newcommand{\fpr}{\texttt{f}}
\newcommand{\cipher}{\mathbf{C}}

\newcommand{\challenger}{\mathcal{C}}
\newcommand{\adversary}{\mathcal{A}}
\newcommand{\rop}{\mathtt{R}}
\newcommand{\public}{\mathtt{P}}
\newcommand{\secret}{\mathtt{S}}

\newcommand{\dis}{\mathsf{d}}
\newcommand{\sdis}{\mathsf{D}}
\newcommand{\s}{\mathsf{S}}

\newcommand{\vr}{\mathbf{r}}

\newcommand{\vx}{\mathbf{x}}
\newcommand{\vy}{\mathbf{y}}
\newcommand{\fx}{\mathbf{X}}
\newcommand{\fy}{\mathbf{Y}}

\renewcommand{\paragraph}[1]{\smallskip \noindent {\textbf{#1}}}

\newtheorem{theorem}{Theorem}

\begin{document}

\title{Outsource Photo Sharing and Searching for Mobile Devices With Privacy Protection}

\author{\authorblockN{Lan Zhang\authorrefmark{1},
Taeho Jung \authorrefmark{2},
Cihang Liu \authorrefmark{1},
Xuan Ding \authorrefmark{1},
Xiang-Yang Li\authorrefmark{2},
Yunhao Liu\authorrefmark{1}}
\authorblockA{\authorrefmark{1} School of Software, Tsinghua University}
\authorblockA{\authorrefmark{2} Department of Computer Science, Illinois Institute of Technology}
}



\maketitle

\begin{abstract}
With the proliferation of mobile devices,
 cloud-based photo sharing and searching services are becoming common
 due to the mobile devices' resource constrains.
Meanwhile, there is also increasing concern about privacy in photos.
In this work, we present a framework \ourprotocolNSP,
  which enables cloud servers to provide
  privacy-preserving photo sharing and search as a service to mobile device
  users.
Privacy-seeking users can share their photos via our framework to allow only their
 authorized friends to browse and search their photos using resource-bounded
 mobile devices.
This is achieved by our carefully designed architecture and novel outsourced privacy-preserving computation protocols,
 through which no information about the outsourced photos or even the search contents (including the
 results) would be revealed to the cloud servers.
Our framework is compatible with most of the existing image search
 technologies, and it requires few changes to the existing cloud systems.
The evaluation of our prototype system with 31,772 real-life images shows the communication and computation efficiency of our system.
\end{abstract}

\vspace{-0.08in}
\section{Introduction}
\vspace{-0.05in}
\label{sec:introduction}

With the increasing population of smart personal devices (\eg,
smartphone, tablet PC)  as well as the emergence  of wearable devices
(\eg, Google Glass),
  huge amount of photos are produced everyday.
The data volume of photos are growing exponentially due to the
 high-resolution on-board cameras, and this makes
 the photo management and sharing challenging to mobile devices.
Facing such challenge, users often choose to outsource the burdensome image storage and searching
 to cloud servers such as Amazon Cloud Drive, Dropbox and some image-oriented cloud (Cloudinary).
Various social networking systems (Flickr, Facebook, Google Plus \etc) also provide photo sharing services for personal uses.


However, privacy becomes a critical issue when photos are outsourced to third parties.
For example, the image recognition technique introduced by Facebook was very controversial in 2011,
 because the objects in users' photos such as faces and cars can be automatically recognized \cite{lowe2004distinctive,bay2008speeded}.
 And tracking and stalking become easier with various image search engines (\eg, Google Image Search, Yahoo! Image Search).
This controversy finally made Facebook to switch off its face recognition service in 2012.
But it has brought back the functionality recently due to the need for image search.
To handle such privacy issues, Google has decided to forbid the face recognition functionality on Google Glasses.
Many users are still struggling to disable face recognition in Facebook.
Simple privacy policy enforced by some companies cannot guarantee the privacy protection,
 especially when the search can be automatically conducted by a machine.
Considering the
rich sensitive information (\eg, people, location and event) embedded in the
photos, privacy in images is in urgent need of protection.

To some extent, the above privacy concerns come from the fear that our
photos might be illegally searched by a malicious hacker, and this is
probably one of the primary reasons why users want to get rid of face recognition. 
However, the object recognition techniques could bring powerful
ability to image search, and we believe simply disabling automatic
recognition is not the best solution to the privacy problem
because it also eliminates the potential utilities lying in the image
search functionality. Ideally, privacy-sensitive users should have an option to use the secure version of photo
sharing and searching system with little extra overhead, in which image search with
object recognition is allowed for authorized users but the privacy leakages due to
automatic recognition are prevented.

To achieve this vision,
 in this paper, we first design a framework \ourprotocolNSP,
 allowing mobile device users to enjoy photo sharing with fine-grained privacy protection policies,
 which can be provided by any cloud to attract privacy-seeking users.
 Via our framework, an \textit{owner} can share his photos in a cloud without unintended access to his photos,
 and an authorized \textit{querier} can send photo queries to conduct image search on others' photos.
 To allow resource-bounded mobile devices to search a huge volume of photos,
 \ourprotocolNSP~outsources the heaviest computation and storage tasks to the cloud,
 and only the private parts of photos are selectively protected to further reduce the computation and communication overhead.
 Despite such outsourcing, \ourprotocolNSP~does not reveal the private image contents or the query contents (including its result) to the cloud.
 In the later section, to more aggressively enhance the performance, 
 we further introduce the optimized $\ourprotocolNSP_{bin}$ where the computation overhead is reduced by half
 with only a little loss of accuracy.

\CUTTaeho{
One cannot directly rely on existing techniques to design such services due to the following reasons.
Various works propose well-designed solutions
 for the large-scale image search \cite{sivic2003video,jegou2007contextual}, but
 most of them focus on the performance aspect rather than the privacy aspect.
Searchable symmetric encryption is proposed to
 enable search over encrypted text documents \cite{abdalla2005searchable,bellare2007deterministic,reza2004sse}, but they are designed for the text documents, and the performance is poor when applied to complex high-resolution images.
Existing privacy protection schemes for image management (\eg, P3 \cite{ra2013p3}) focus on the storage aspect, and the system does not meet the growing needs for the image search.
In GigaSight \cite{gigasight-mobisys13}, based on time, location, and
content, privacy sensitive
information is  removed from the video, but it does not
consider the privacy preserving search as well.
}

\CUTXY{
\noindent \textbf{Challenge 1}: Various works propose well-designed solutions
 for the large-scale image search \cite{sivic2003video,jegou2007contextual}, but
 most of them focus on the performance aspect other than the privacy aspect.

\noindent \textbf{Challenge 2}: Searchable symmetric encryption (SSE) is proposed to
 enable search over encrypted text documents \cite{abdalla2005searchable,bellare2007deterministic,reza2004sse}. They are designed for the text documents, and the performance is poor when applied to complex high-resolution images.

\noindent \textbf{Challenge 3}: Exiting image privacy protection schemes (\eg, P3 \cite{ra2013p3}) design privacy preserving image storage systems, but they do not consider efficient search and cannot meet the growing needs for the image search.

}


\ourprotocolNSP~ can be considered as a
step towards easily deployable frameworks for privacy-preserving photo
sharing and searching services among mobile device users, taking advantage of the
availability of cloud servers who possess powerful computation and
storage abilities.

In summary, our contributions can be summarized as:
\begin{enumerate}[1.]
\item We propose a novel modularized photo sharing and searching framework to let mobile device users outsource photos
      and the majority of heavy jobs (storage, access control and searching) to cloud servers, without breaching users' photo-related privacy.
\item We design two outsouced vector distance computation protocols for both real and binary vectors,
 which are the core of the framework.
      Different from the existing multi-party computation based methods, our protocols enable efficient vector distance computation
      in a non-interactive way, which means the photo owner does not have to interact with the cloud or the querier.
\item Our framework is compatible with common image services and feature based image search methods.
 To achieve nice user experience, all privacy protection modules work automatically and are transparent to users.
 After the privacy setting, users can enjoy the sharing and searching services as usual.
We implement and evaluate our framework using 31,772 real-life images
 on both smartphones and laptops. The evaluation shows that very low extra overhead is incurred by our method.
\end{enumerate}

\vspace{-0.08in}
\section{Backgrounds and Motivation}
\label{sec:background}
One of our main contributions is enabling efficient photo sharing and
searching on encrypted photos.
In order to achieve high search accuracy,
 \ourprotocolNSP~leverages the state-of-the-art image search technologies in computer vision field.
Here, we briefly review the search techniques, and then discuss the privacy issues emerging from them in this section.


\vspace{-0.05in}
\subsection{Descriptors Based Image Search}\label{section:image_descriptor}
\vspace{-0.05in}


Images are usually searched by their contents.
Different types of \textit{visual descriptors} are proposed to model the visual characteristics of the image,
\eg, color, intensity, texture or objects within the image.
Various image contents can be recognized and localized (\eg, people\cite{leibe2005pedestrian} and face\cite{viola2004robust}) using visual descriptors.
Among these, human face detection received extraordinary attention and is one of the most mature object detection techniques so far\cite{turk1991eigenfaces,viola2004robust}.


The feature descriptor is usually constructed as a set of numeric vectors, denoted as \textit{feature vectors}.
There are some statistical feature vectors (\eg, intensity/color histograms).
Also, many well designed visual descriptors are proposed to achieve accurate image search, \eg,
 SIFT \cite{lowe2004distinctive} and SURF\cite{bay2008speeded}.
In those works, each feature vector is generated from
 an interest point of the image to describe the visual characteristics
 around the point.
Interest points are pixels containing distinguishing information of the image \cite{mikolajczyk2004scale}.
In general,
all feature vectors belonging to the same descriptor have the same
dimension (\eg, SIFT has 128 dimensions and SUFR has 64
dimensions).
The numeric type of vectors may be real number \cite{lowe2004distinctive,bay2008speeded}
 or binary \cite{calonder2010brief,Peker-2011},
 and different types of vectors are used for different applications.
Specifically, with a little accuracy loss,
 binary descriptors are usually more efficient in computation and suitable for resource-restricted
 mobile applications.
We design our framework capable of dealing with both real number and binary feature vectors.


\CUTTaeho{
\subsubsection{Conceptual Descriptor}
Besides,
 conceptual descriptors are extensively used in image search engines
 to describe the conceptual content of images,
 \eg, the metadata, tags and relevant texts.
The metadata may contain location stamp, time stamp of the image; the tags and relevant texts may describe the identity, objects and actions in the image.
For search task, conceptual descriptors are usually represented by numeric vectors too.\bigskip
}



Given a query image, one needs following three steps to search the top-$k$ similar images from the database. Firstly, pre-defined image descriptor is extracted from the query image. Secondly, each feature vector in the query image is compared with feature vectors from the database images. Thirdly, similarity score for every database image is measured based on the vector comparison and finally the top-$k$ high-score images are returned to the querier.
%
%


\CUTTaeho{
\begin{figure}[h]
\begin{center}
\includegraphics[width=0.85\linewidth, clip,keepaspectratio]{matching.eps}\vspace{-5pt}
\caption{Example of interest points \& matching}\vspace{-10pt}
\label{fig:matching}
\end{center}
\end{figure}

Figure~\ref{fig:matching} shows an example of interest points
 detected by Hessian-based detectors \cite{lindeberg1998feature} as well as the matching result when using SURF descriptor \cite{bay2008speeded}.
}

\CUTTaeho{
\subsubsection{Search Matching Feature Vector}
The majority computation overhead comes from finding the top-2 feature vectors from a queried descriptor
 for each vector in the querying descriptor.
For some well modeled objects, very limited feature vectors
 are sufficient for a descriptor,
 \eg, for face recognition algorithm, $9$ feature vectors can distinguish different faces.
In this case a simple linear search is efficient for nearest neighbor search.
For some complicated images with hundreds of feature vectors,
 some strategies are proposed to speed up the search.
K-dimensional tree (k-d tree) based fast approximate nearest neighbors algorithms \cite{muja2009fast},
 are extensively used in the computer vision field,
 which can speed the matching of high-dimensional vectors
 by up to several orders of magnitude compared to linear search
 with little accuracy lose.

The original k-d tree based method splits the data in half at each level of the tree on the dimension
for which the data exhibits the greatest variance.
When searching the trees, a single priority queue
is maintained across all the randomized trees so that
search can be ordered by increasing distance to each
bin boundary.

\subsection{Visual words based image search}

Most recently,
 visual words are used by more work for large scale image search.
The space of descriptors is quantized
 to obtain the visual vocabulary (this process is performed off-line),
 which reduces the cardinality of the vector space. Then, value of each di
An image is then represented by the
frequency histogram of visual words obtained by assigning
each descriptor of the image to the closest visual word.
Then image indexing and search are based on the similarity of histogram.
}

\vspace{-0.05in}
\subsection{Privacy Implications}
\vspace{-0.05in}

Rich content of photos raises various privacy implications.
There are many mature techniques to detect and recognize the objects within the photos as
aforementioned.
These techniques can possibly be used to automatically
analyze the photos to mine sensitive information with various data
mining techniques.
Combining the location stamps and time stamps embedded in a photo, more
sensitive information about the person may be derived
(\eg, home location, occupation, level of incoming).
Therefore, the private part (denoted as Region Of
Privacy (\textit{ROP}) hereafter) of a photo needs to be protected,
so that no human or machine runnable algorithm can learn sensitive
information in the photo.

Besides the outsourced photos, the query sent to the cloud side incurs
privacy implications as well. Even though the uploaded photos are well
protected via encryption so that the cloud does not gain useful
information of them, their contents can be easily deduced if the
queries' contents and results are revealed to the cloud.
Since the entire search process should be outsourced to the cloud for resource
saving, protecting queries' contents as well as the results
is equally important to protecting the uploaded photos.


\vspace{-0.08in}
\section{System Overview}
\label{sec:system}
\begin{figure}[h]
  \centering
  \subfigure[Photo Owner Side]{\label{fig:system_overview-a}
    \includegraphics[width=0.8\linewidth, clip,keepaspectratio]{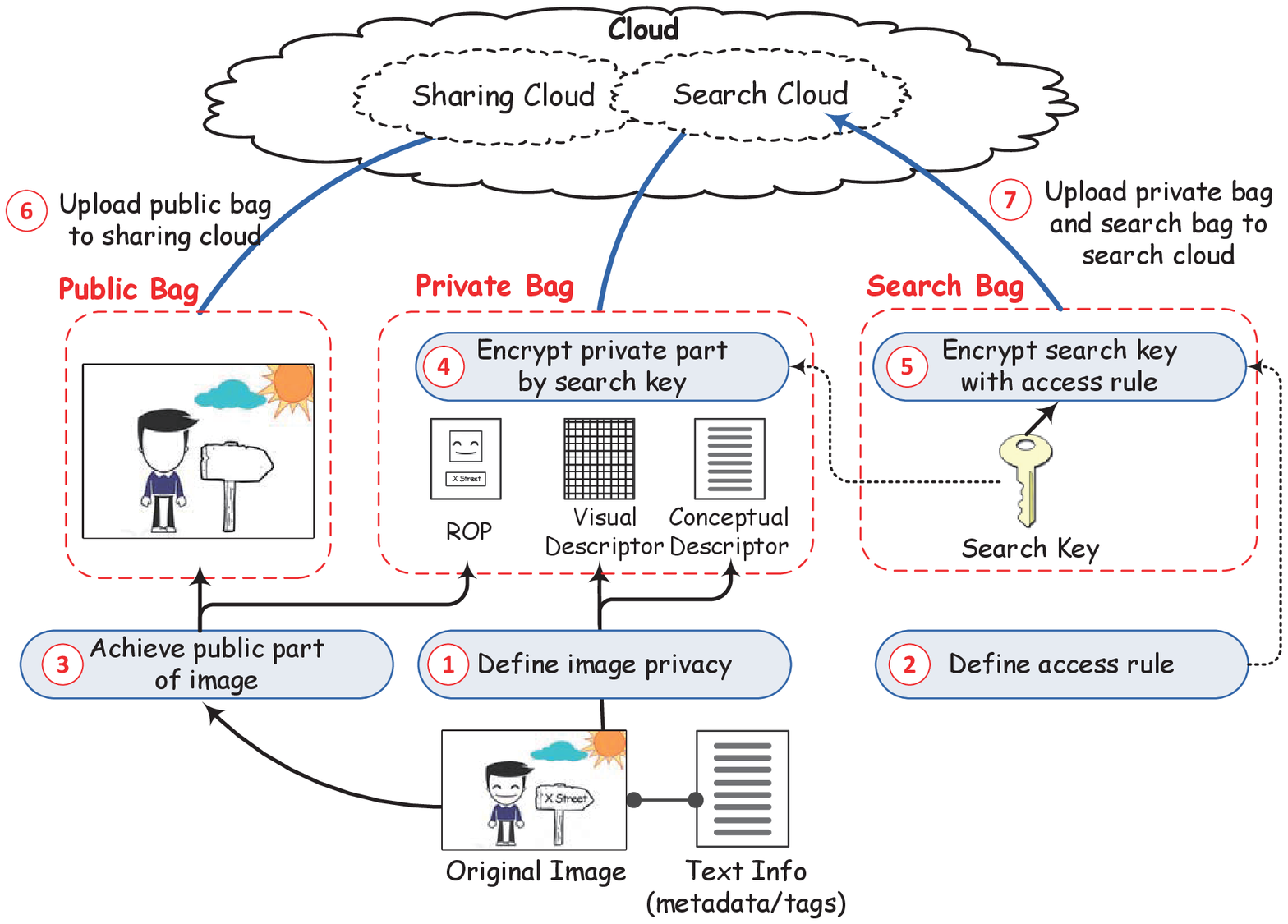}}
  \subfigure[Querier Side]{\label{fig:system_overview-b}
   \includegraphics[width=0.8\linewidth, clip,keepaspectratio]{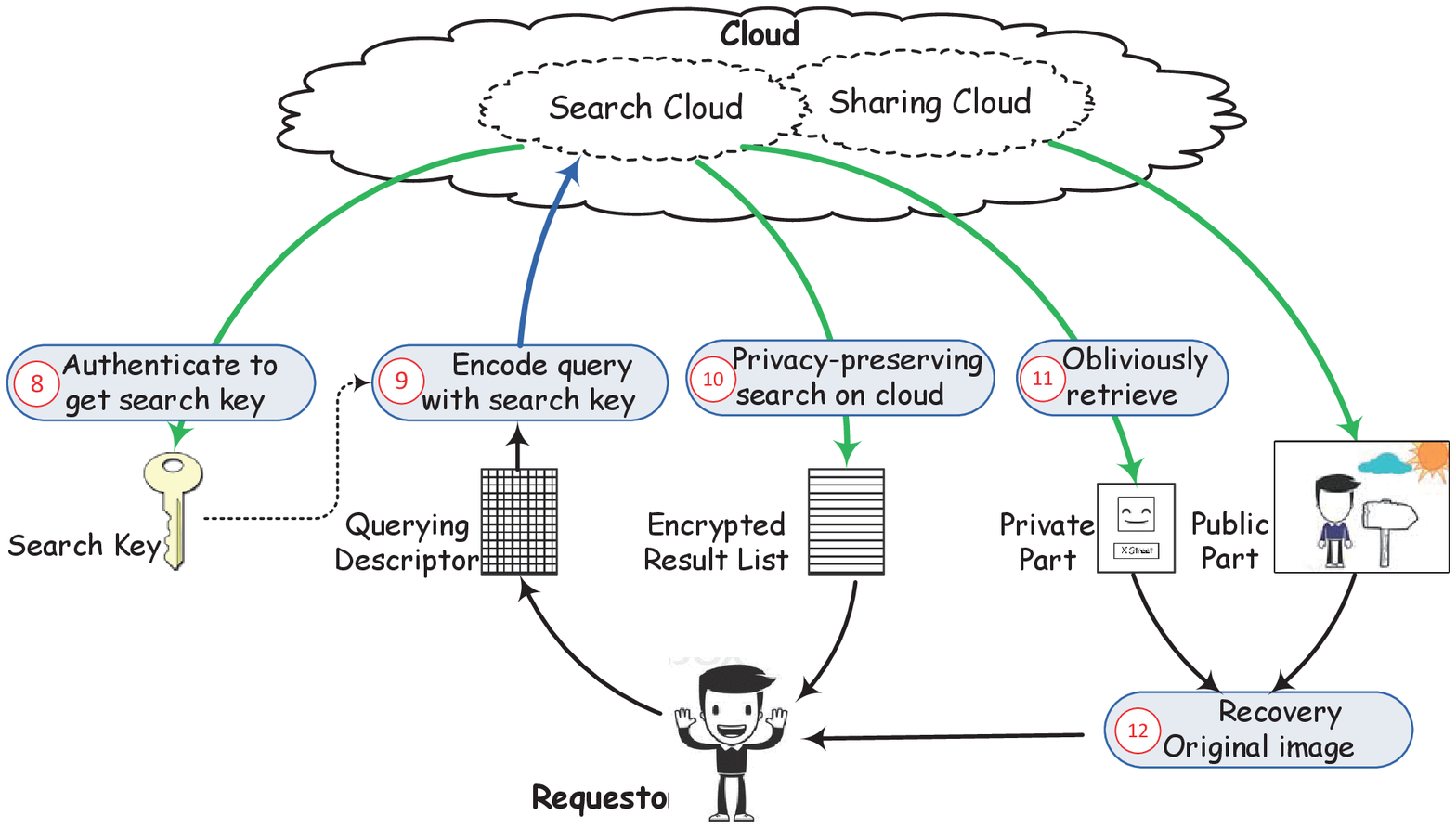}}\vspace{-5pt}
  \caption{\ourprotocol System Overview}\vspace{-10pt}
  \label{fig:system_overview}
\end{figure}

\ourprotocolNSP~is a novel framework allowing clouds to provide privacy preserving photo sharing and searching services to the mobile devices users.
It can attract users who need both outsourced photo service and privacy protection.
Figure~\ref{fig:system_overview} illustrates the architecture and workflow of our framework\footnote[2]{We
  logically divide the cloud into \textit{sharing cloud} and
  \textit{search cloud} for explanation purpose, but revealing
  this structure does not breach users' privacy at
  all.}. With this framework design, \ourprotocolNSP~ can provide the following services:
  (1)privacy-preserving photo storage outsourcing;
  (2)fine-grained photo sharing with privacy protection enforcement;
  (3)light-weight photo searching for mobile devices.


%



\vspace{-0.05in}
\subsection{Privacy Preserving Photo Storage}\label{sec:privacy_storage}
\vspace{-0.05in}

Before users upload their photos to cloud
servers for sharing, the photos need to be pre-processed.
Firstly, the region of privacy (ROP), which is a rectangle defined by two pixel-level
coordinates (top-left and bottom-right) on the photo, is either automatically or manually determined.
In the automatic manner, the user can define a category of objects as private content, \eg, faces.
Then all private objects will be detected by object recognition algorithm and set as ROPs, \eg, the face in Fig.~\ref{fig:mask}.
Otherwise, the owner can also manually define the ROP by selecting a rectangle region on the photo.
Then, the feature vectors of the ROP are extracted according to the
definition of the image descriptor (Section \ref{section:image_descriptor}).
Note that, hereafter we use the human faces as example ROPs of photos in this work,
 but other objects such as pedestrians and cars, can also be defined as ROPs with
corresponding recognition algorithms.
Moreover, except defining ROP, all the following operations are conducted automatically
by the system and transparent to users.

After the ROP is selected, it is separated into public part and secret
part, where the public part doesn't contain any sensitive information
and the secret part is encrypted such that only the authorized
users with keys can access to it and recover the original ROP. We review
the following three different methods for the separation:

\noindent 1. \textbf{Mask}: fills public part of ROP with solid black (all intensity values `0') and takes the original ROP as secret part.

\noindent 2. \textbf{P3\cite{ra2013p3}}: separates ROP based on a threshold in the DCT frequency domain; sets the higher frequency part as secret part and the remaining as public part.

\noindent 3. \textbf{Blur\cite{mcdonnell1981box}}: a normalization box filter is applied to ROP to generate the public part; subtracts the public part from ROP in a pixel-wise way to achieve the secret part.

\noindent
Then, the public part of the whole photo is produced by replacing its ROP with the public part of ROP (as shown in Fig.~\ref{fig:mask}).
Our experiment (Section~\ref{sec:eva}) shows that all three methods
are resistant to automatic detection algorithms, but the blur based
method outperforms others in the storage cost, hence we adopt the blur
as the default separation method in \ourprotocolNSP. After extracting the
private part from ROP, the owner encrypts the secret part as well as
its image descriptor as a \textit{private bag}, and uploads the private
bag to the search cloud. Then, he also uploads the public part of the
original photo as a \textit{public bag} to the sharing cloud (Fig.~\ref{fig:system_overview-a}).

\CUTTaeho{
\begin{figure}[h]
\begin{center}
\includegraphics[width=0.8\linewidth, clip,keepaspectratio]{checkbox.eps}\vspace{-5pt}
\caption{Hierarchical Automatic ROP Detection}\vspace{-10pt}
\label{fig:checklist}
\end{center}
\end{figure}
}

\vspace{-0.05in}
\subsection{Fine-grained Photo Sharing}
\vspace{-0.05in}

\ourprotocolNSP~allows fine-grained photo sharing among users. The photo owner uses an access control scheme (\eg, \cite{bethencourt2007ciphertext, chase2009improving}) to encrypt the \textit{search keys} so that only the authorized users with certain attributes can obtain search keys.
As the Step 5 in Fig.~\ref{fig:system_overview-a}, the owner encrypts the search keys under the \textit{access rule} that he defines,
 and the encrypted search keys are uploaded to the sharing cloud and made published.
Obtaining the search key, the authorized user can generate valid photo queries and decrypt the private part of ROP.
The completed original images can be recovered simply by merging the public parts of images and the private parts of ROPs.
Here, all these operations are also automatic and transparent to users, and the authorized user can browse the shared images as usual.

\vspace{-0.05in}
\subsection{Light-weight Photo Searching}
\vspace{-0.05in}
When a querier wants to search a photo among someone else's photos, he
pre-processes the querying photo to achieve the corresponding image
descriptor. Then, only if satisfying the owner's access rule, he can
retrieve the search keys to search on the owner's photos, but it is
the cloud who conducts the searching job and returns the result to the
querier obliviously, \ie without knowing contents of the owner's ROPs
or the contents of the query photo. After fetching the query result,
as mentioned above, system generate the original image for the querier transparently.
For the querier, the whole system appears like common image search systems.
Fig.~\ref{fig:system_overview-b} illustrates the search procedure.

\begin{figure}[t!]
\begin{center}
\includegraphics[width=0.95\linewidth, clip,keepaspectratio]{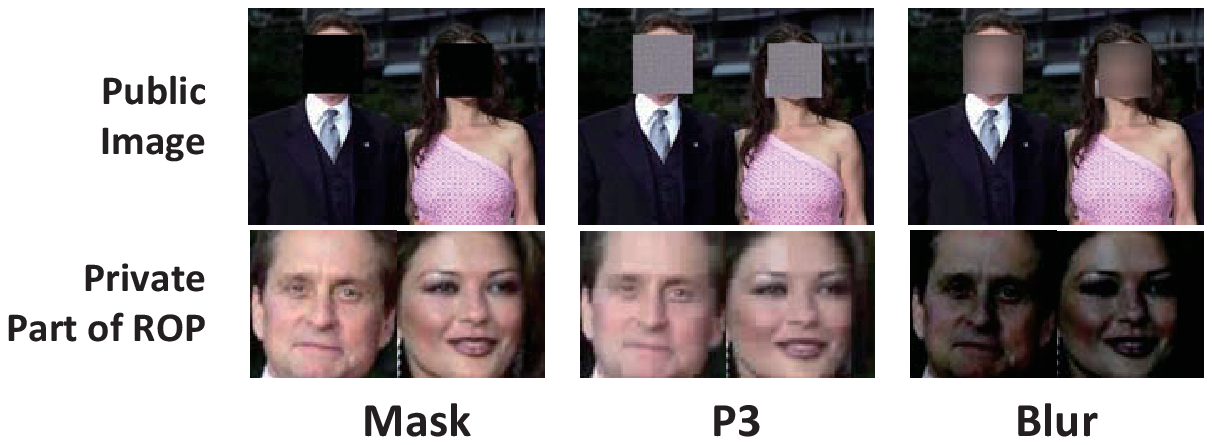}\vspace{-5pt}
\caption{Public/Private Part of ROP. Images in the upper row are public part of image and images in the lower row are private part of ROP.}\vspace{-20pt}
\label{fig:mask}
\end{center}
\end{figure}

\vspace{-0.05in}
\subsection{System Design Goals}
\vspace{-0.05in}


Our system is designed to achieve efficiency, privacy protection and accuracy goals.

$\bullet$ \textbf{Efficiency}:
To overcome the resource limitation of mobile client,
 operations at the user side should be light-weight,
 and most of the expensive computations should be outsourced to the cloud side.

$\bullet$ \textbf{Privacy Preservation}:
Users outsource not only the storage of photos but also the searching to the cloud side in \ourprotocolNSP.
Therefore, the framework is expected to protect users' privacy in various aspects:

 1. ROP Privacy: Unauthorized party should not learn secret part of ROP including cloud servers,

 2. Query Privacy: Cloud servers should not learn query photos,

 3. Result Privacy: Cloud servers should not learn search results,
\noindent which are all non-trivial challenges since cloud servers are the party who conducts the searching jobs on the photos stored at his side.

$\bullet$ \textbf{Accuracy}:
Introducing the privacy protection mechanism should not bring much accuracy loss. That is, the search result from \ourprotocolNSP~should be comparable with traditional image search technologies conducted on plain texts of photos.

\vspace{-0.05in}
\subsection{Threat Model}
\vspace{-0.05in}

\textit{W.l.o.g.}, we assume \textbf{curious-but-honest} cloud servers and \textbf{malicious} users in this work. Cloud servers will follow the protocol specification in general, but they will try their best to harvest any information about user's photos. This is a justifiable assumption because deviating from the protocol and not returning a correct search result will lead to bad user experience as well as potential revenue loss of the service provider. However, they might conduct extra work to illegally harvest useful information from the protocol communications in order to infer the \textbf{secret part of ROP} or the \textbf{contents of queries}, which is sensitive information to be protected. On the other hand, queriers may misbehave throughout the protocol to infer the \textbf{search keys} to forge a valid photo query, where the search keys are supposed to be kept secret as well.

\CUTTaeho{
Our system is designed to support different
 levels of privacy-preserving image search:
\begin{compactenum}
\item Image level search: a user inputs a required image,
 the system returns the image or a set of images most likely
 containing this sub-image.
This is the basic and most general image search.
In this level of image search,
 the system doesn't care the content of the required image.
So the search is based on the image similarity.
In our system, the similarity can be measured by
 visual descriptors or visual words.

\item Object level search: To search the images of a specific object,
 a user inputs an image of the required object,
 the system return images containing this object.
 In most applications,
 there are many images for a single object, \eg, a specific person, a logo, a landmark.
For the system implementation,
 there is a template descriptor for a known object.
Compared with the image level search,
 a modification is required for the data storage.

\item Conceptual search:User input the search keywords (text description of the target image),
 \eg, the person name or color name.
 and the related images are returned.
\end{compactenum}

Though the image search itself is a difficult
 issue, the privacy consideration greatly increases the challenges.
Facing these challenges,
 we designed a series of methods to achieve the following privacy requirements
 during the storage, sharing and search processes:
 \begin{compactenum}
 \item the private regions of images can only be accessed (viewed/detected) by the owner and valid visitors;
 \item protecting the owner's and requestor's visual and conceptual descriptors from untrusted parties;
 \item protecting the similarities between descriptors from untrusted parties.
 \end{compactenum}
Besides, our system also protects requestors' attributes
 and retrieve patterns, \ie the search process must be oblivious to the services provider and other parties.

For the system performance,
 we design our system to achieve comparable accuracy with exiting
   mechanisms without privacy consideration,
   meanwhile, not to greatly increase the storage, communication and computation cost.
Moreover, the ease of use, multiple image formats, the extensibility and scalability
 should also be taken into consideration during the system design.

\subsection{User Defined Privacy Protection Policy}
The image privacy protection policy is set by the image owner in the client side.
As shown in Fig.~\ref{fig:system_overview},
 two aspects of protection policy are determined:
 in Step {\large{\textcircled{\small{1}}}}, the private content is defined;
 and in Step {\large{\textcircled{\small{2}}}} the access rule is defined.

\subsubsection{Private Content}
In most cases,
 the private content (\eg faces and landmarks) in an image are subregions of the image.
For an image $I$,
 we define a private subregion as a \emph{Region Of Privacy} (ROP), denoted as $\rop=(tl, br)$,
 here $tl$ and $br$  are the pixel-level coordinates of its top-left and bottom-right in $I$.
Let $\rop(I)$ denote the sub-image extracted from $I$ according to $\rop$.
In our system,
 both automatic and manual ways are supported for an image owner to
 determine the ROP.
For the ease of use,
 our system provides a hierarchical check list for the owner to define
 private content (\eg, an object or a class of objects, recognized by
 the state-of-the-art computer vision technologies, or tagged by user).
For object classes with mature automatic detection methods,
 the owner can select which class is private, \eg, people faces.
For the known objects with tags, \eg people faces tagged with names,
  the owner can select the specific private objects by checking the name.
Specially, when the private object is a whole image,
 the owner just need to select the ID of this image.
Fig.~\ref{fig:checklist} shows an example check list of our system.
From the checked items,
 our system runs the object detection algorithm over the input images,
 and records the ROPs of each images using  bounding boxes.
 as the red rectangle in Fig.~\ref{fig:checklist}.

So far, most privacy related objects,
 \eg faces and texts, can be automatically detected.
In case  there are still some sensitive objects or scenes
 cannot be automatically detected,
 a manual ROP definition mode is also supported.
In this mode,
 user defines the ROP $\rop$ by selecting its bounding box $(tl, br)$.

While the ROP $\rop_{I_x}$ of an image $I_x$ is determined,
 for the vision feature based search,
 the client of our system will generate a feature descriptor $\fx$ for
 this sub-image $\rop_{I_x}(I_x)$.
First, a set of interest points $p=\{p_1, p_2, \cdots, p_\alpha\}$
 are detected with a feature detector.
Then, a fix-length feature vector $\vx_i$ will be generated from
 the interest point $p_i$ of this sub-image using a describing method, \eg, SIFT \cite{lowe2004distinctive} or SURF.
These feature vectors compose the feature descriptor $\fx=\{\vx_1, \vx_2, \cdots, \vx_\alpha\}$.

\subsubsection{Access Rule}

To facilitate the cloud-based image sharing
 while protect the image privacy from unauthorized users,
 the owner makes access rules for the ROPs to determine how he wants to share his images.
In most social networking and image sharing systems, \eg, Facebook, Flickr,
 each user is assigned with a set of attributes, \eg, school, profession.
The access rule is defined by determining the
 attributes of the authorized visitors.
Specifically, let $\mathbb{A} = \{a_1, a_2,\cdots, a_m\}$ be the universal set of attributes.
The owner need  define an access structure, often expressed by a tree structure
 with AND and OR gates, \eg, \cite{bethencourt2007ciphertext}.
The owner-defined access structures over $\mathbf{A}$ will be used in
 Step {\large{\textcircled{\small{5}}}} of the privacy concealing component
 to enable access control of image privacy.

Before uploading to cloud,
 all data will be packed into two bags automatically by the client side of our system:
A \emph{public bag} containing the public part of the images without any sensitive information,
 which can be uploaded to any exiting sharing or storage cloud;
A \emph{private bag} containing the encrypted sensitive information
 for search, image recovery and access control.
In the following two subsections,
 we will present the mechanisms of privacy concealing and encryption
 of these two components.

\subsection{Public Image Privacy Concealing}
\label{sec:concealing}

In Step {\large{\textcircled{\small{3}}}}, for each image $I_x$,
 sensitive information of its ROP
 will be concealed from the original images to generate a public image $\public(I)$.
The public image  can prevent unauthorized users
 and automatic feature detection by machine.
As a result, the public image can be used for common image sharing systems.
The secret images are required when recovering the original image.
A secret image $\secret(\rop(I_x))$ is then generated for
  the sensitive information from the ROP.
Fig.~\ref{fig:mask} illustrates the public images and secret images
 generated by different concealing methods.

Given $\rop(I_x)$, which is the sub-image of $I_x$ extracted according to ROP definition,
 there are several optional methods to conceal
 the sensitive information to produce a public part $\public(\rop(I_x))$.
For each processing method,
 a corresponding secret part $\secret(\rop(I_X))$ is generated.
\begin{enumerate}
\item \emph{Mask} based method: the public part $\public(\rop(I_X))$ is filled with solid black (all intensity values equal `0'); the secret part is the same as the original sub-image,
\item \emph{Threshold} based method: a threshold is applied to $\rop(I_x)$ to
  separate it in a defined domain (\eg P3 separates image by threshold
  in the frequency domain, bit plane separates image by extracting
  specific bits from intensity values); $\public(\rop(I_x))$ contains
  the less significant information, while $\secret(\rop(I_x))$
  contains the more significant information;
\item \emph{Blur} based method: a normalization box filter is applied to
  $\rop(I_x)$ to generate $\public(\rop(I_x))$; the private part is
  generated by subtracting the blurred public part from original
  sub-image.
\end{enumerate}
Except P3 which requires a DCT processing,
 for the other methods the secret part can be generated by
 subtracting the public part from the original sub-image pixel-wise,
 \ie $\secret(\rop(I_x))=\rop(I_x)-\public(\rop(I_x))=\rop(I_x)$.
Then the public part of the  original image $I$
 is produced by simply replacing its ROP with the public part
 of ROP, \ie $\public(I_x) = I_x - \rop(I_x) + \public(\rop(I_x))$.
Fig.~\ref{fig:mask} shows an example of privacy concealed result of the public part
 of image and their corresponding secret parts of the ROP.
The three methods are all resistant to automatic detection and human recognition.
As we will present in our evaluation ,
 the blur based method performs best in both storage cost and aesthetic measure.
Hence, \ourprotocol uses blur as the default method
 for public part privacy concealing.

\subsection{Private Part Encryption}

For an image $I$, 
 the private part consists of the secret part of ROP, \ie $\secret(\rop(I))$,
 and the feature descriptors $\fx$ of each ROP.
Both $\secret(\rop(I))$ and $\fx$ require protection.
The vision-based search are conducted with feature descriptors.
The great challenge comes from the
 conflicted requirements of privacy protection and efficient search.
In our work, we design a special encryption method for the feature descriptor,
 to provide efficient various privacy-preserving image searches.
In our design, PPS3  outsources as much computation as possible
 to the  cloud with as few interactions as possible with the
 requestor.
And we will present the detail of private part encryption in Section~\ref{sec:search}.

In Step {\large{\textcircled{\small{4}}}},
 after encrypting the feature vectors for an image $I$,
 the owner encrypts the secret part $\secret(\rop(I))$  by his
 public key $PK$.
In Step {\large{\textcircled{\small{5}}}} the private key $SK$ and search keys
 are encrypted by a CP-ABE algorithm \cite{bethencourt2007ciphertext, waters2011ciphertext}.
Let $c = \abe(m)$ be the ciphertext of a message $m$ with the public key $MPK$
 and the access structure $\mathbf{A}$ using CP-ABE algorithm.
}

\CUTXY{
The CP-ABE mechanism mainly consists four components:
\begin{enumerate}
\item Setup: the authority outputs a public parameter $MPK$ and a master secret key $MSK$.
\item Encryption: the algorithm $c = \abe(m)$ encrypts a message $m$ with the public parameter $MPK$ and the access structure $\mathbf{A}$ and produce a ciphertext $c$ that only a user whose attributes satisfy $\mathbf{A}$ is able to decrypt it.
\item Key generation: the algorithm $ASK = \kg(MSK,A)$ takes as input the master secret key $MSK$ and a set $A$ of attributes and produces
 a private attributes-key $ASK$ for $A$.
\item Decryption: the algorithm $\dabe(c)$ decrypts the ciphertext $c$ with the public parameter $MPK$ and the private key $ASK$
 and produce the plain message if the private key's corresponding attributes satisfy the access structure $\mathbf{A}$.
\end{enumerate}
}

\CUTTaeho{
Specifically, our system leverage the CP-ABE mechanism in the following way:
 the owner uses the encryption algorithm to generate $\abe(SK)$, $\abe(\{K^0, K^1\})$
  using the access structure $\mathbf{A}$.
And the encrypted keys will be uploaded to the searching server.
A requestor who wants to search images in the cloud
 first goes to the authority to generate his private decryption key $ASK = \kg(MSK,A)$
 with his attributes $A$.
Then the requestor visits the searching cloud to obtain $\abe(SK)$ and $\abe(\{K^0, K^1\})$
 and tries to decrypt them with his key $ASK$.
If he is an valid user with authorized attributes,
 he can get the correct $SK$ and $\{K^0, K^1\}$
 to enable  further search.
Note that, the Paillier public key $PK$ of the owner is  encrypted by CP-ABE.
The search server will be allowed and only allowed to access the public key $PK$
 for the privacy-preserving cloud-based search.

After Step {\large{\textcircled{\small{5}}}},
 the privacy information in the image have been concealed,
 and all images are packed into two bags:
 the \emph{public bag} consisting of public parts $\public(I_x)$ of all images$I_x$;
 the \emph{private bag} consisting of Paillier encrypted secret parts of ROP $\en_{PK}(\secret(\rop(I_x)))$,
 garbled circuit encrypted feature descriptors $\garbled(\fx)$,
 and attribute based encrypted $\abe(SK)$ and $\abe(\{K^0, K^1\})$.
For image with metadata (\eg, location) or tags,
 each text entry is encoded into a text vector
 using some well-studied method, \eg hashing or cosine vector.
These text vector is also encrypted by the garbled circuit method
 and then included in the private bag to enable privacy-preserving text-based image search.
With the user defined access structure and
 and the CP-ABE encryption,
 \ourprotocol protect the content of private bag
 from \emph{unauthorized ROP access}.

\subsection{Cloud Storage}

The public bag can be uploaded to any common share or storage platform.
Each image has a unique ID and a URL pointing to its public part, here
 the URL is also encrypted by the owner using $PK$.
While the private bag will be uploaded to the cloud server supporting
 our search mechanism.
Fig.~\ref{fig:storage} sketches the storage structure of the private
data on the search cloud.

\begin{figure}[h]
\begin{center}
\includegraphics[width=0.9\linewidth, clip,keepaspectratio]{cloudstorage.eps}
\caption{The private data stored on the search server.}
\label{fig:storage}
\end{center}
\end{figure}

For image owner,
 the search server maintains the pair of Paillier keys and search keys
 encrypted by ABE (an owner could have multiple pairs of Paillier
 keys, each for a group of images that need individual attribute-based
 access control).
Each image could have more than one ROPs.
For each ROP, the server maintains its secret part encrypted by the
 public key $PK$
 and its feature vectors encrypted by the modified garbled circuit.

\subsection{Privacy Preserving Image Search}
When a requestor $P_1$ needs to search an image of a specific owner,
 he obtains the owner's attribute encrypted private key and search key from the server
 and tries to decrypt them with his attributes as described in the previous Subsection~\ref{sec:ppe}.
A requestor with authorized attributes set  can get correct $SK$, $K^0$
and $K^1$.
Then he encodes the vectors of his request feature descriptor $\fy$ using $K^0$ and $K^1$ bit by bit,
 \ie $\code(Y) := \{\code(\vy_1), \cdots, \code(\vy_n)\}$.
For each bit $\vy_j(k)$ of $\vy_j$, if $\vy_j(k)=0$, $\code(\vy_j(k))=K^0$, else $\code(\vy_j(k))=K^1$.
Requestor $P_1$ uploads $\code(\fy)$ to the search server $P_a$.
Note that the order of $K^0$ and $K^1$ is revealed to $P_1$,
 but unknown to the server $P_a$, since the server $P_a$ cannot pass the attribute authentication.
So the request descriptor $\fy$ is protected from $P_a$.
Based on our design for feature vector encryption in Subsection~\ref{sec:ppe},
  an efficient \emph{noninteractive} privacy-preserving feature vector distance protocol is presented as Protocol 1.
The first step leverages our garbled circuit encrypted feature vector to compute
 the bitwise encrypted XOR result between $\vx_i(k)$ and $\vy_j(k)$.
The second step leverages Paillier homomorphic addition to get the squared distance between $\vx_i$ and $\vy_j$.


\subsection{Oblivious Image Retrieve with Access Control}

Until now,
 the server and other potential adversaries learn nothing
 about the content of queries and ROPs.
But when users retrieve images,
 statistics about which images they retrieve could also reveal
 some information about images and queries, \eg which images contain
 the most popular person,
 or who may have common friends appearing in same images.
In our system, we use a simple oblivious retrieve method
 to maximize the privacy protection for both image owner and requestor:
 which meets the following requirements
 (1) only the user with authorized attributes can access the corresponding images;
 (2) the search/storage server doesn't learn the attributes of a user when he retrieve images;
 the search/storage server doesn't learn exactly which image a user
 retrieve.

To achieve this requirement,
 an $k$-out-of-$n$ oblivious transfer (OT), \eg,\cite{camenisch2007simulatable}
 is usually conducted combined with some access control policy, \eg,
 \cite{camenisch2007simulatable, camenisch2009oblivious}.
But the existing OT schemes do not hide the privacy information from the database server,
 and the construction is usually expensive.
In our system,
 the server has $N$ image records $\mathbf{R}=\{R_1, \cdots R_N\}$.
Each record $R_i$ contains the URL to its public part
 and information of its private part.
The requestor has a secret  subset $\sigma \subseteq \{1,2 \cdots,N\}$
  which are the indices of his desired images computed using image
  matching method, and $|\sigma| = k$.
To achieve the requestor's $k$-out-of-$n$ selection oblivious to the server,
 the requestor can use $\sigma'$ with randomly added indices to blind the server,
 here $\sigma \subset \sigma'$ and $|\sigma'|=n$.
Note that $n$ is not equal $N$,
 instead, for large dataset with big $N \gg n$.

With the encoded request and the oblivious image retrieval strategy,
 \ourprotocol hides the search content from untrusted parties.
}

\CUTXY{
Obtaining the records from the search server and decrypting
 the URLs and ROPs,
 the requestor retrieves the public parts by URLs.
A malicious search server may
 analyze the network flow of the public part storage server
 and attempt to correlate it with the private part retrievals.
Here we emphasize that,
 even knowing the content of the public part,
 the search server cannot learn its corresponding entry
 in the privacy database,since the URL is encrypted.
So it cannot decide which image entry is retrieved.
With an anonymous communication,
 a server cannot tell who is requesting the public parts,
 and a better protection of the requestor privacy can be achieved.
}

\CUTTaeho{
\subsection{Image Recovery}
Fetching the public and private parts of an image to the client,
 the client need to recovery the original image from both parts.
For different mask mechanism, the recovery method varies.
Except the method P3, which gives a DTC based recovery method,
 for the other privacy concealing methods,
 the recovery is simply adding the decrypted secret part of ROP to the public part of image,
 \ie $I = \public(I) + \secret(\rop(I))$.
}

\vspace{-0.08in}
\section{System  Design}
\label{sec:preliminary}

In this section,
 we first present the building blocks of our system,
 and then give the detail of our non-interactive private image search protocol,
 which is the core of the system and one of our main contributions.

\vspace{-0.05in}
\subsection{Building Blocks of Our System}
\vspace{-0.05in}
\ourprotocolNSP~ is a modularized and well integrated image sharing and searching system, which consists of several building blocks.

\subsubsection{Image Search}

Image search is composed of three steps: image descriptor extraction, finding matching vector and similarity score calculation.

\paragraph{Extracting Image Descriptor.}
As described in Section~\ref{sec:background},
the visual descriptor is extracted from the \textit{interest points} of each photo, where the interest points are automatically detected (\eg, \cite{mikolajczyk2004scale}). Then, the descriptor $\fx=\left\lbrace \vx_1,\vx_2,\cdots \right\rbrace$ of an image $I_x$ is extracted, where $\vx_i$ is a feature vector.

\paragraph{Matching Feature Vector.}
Given a feature vector $\vx\in \fx$ and another descriptor $\fy$,
let $\dis\left(\vx,\vy\right)$ be Euclidean distance between two feature vectors $\vx\in\fx$ and $\vy\in\fy$.
Then given the $\vx$'s nearest neighbor $\vy_{nn}\in\fy$, $\vx$ and $\fy$ are a \textit{matching pair} iff:
\vspace{-0.1in}
\begin{displaymath}\footnotesize
\delta\left(\vx,\fy\right)=\frac{\dis\left(\vx,\vy_{nn}\right)}{\min_{\vy\in\fy-\left\lbrace\vy_{nn}\right\rbrace}\left(\dis\left(\vx,\vy\right)\right)}<\alpha
\end{displaymath}
That is, iff the ratio between nearest distance and the second nearest
distance is less than a threshold $\alpha$, $\vx$ and $\fy$ are a
matching pair. For most object recognition algorithms, $\alpha$ is set
as $0.5$.

\paragraph{Similarity Score.}
Given a querying descriptor $\fx$ and a queried descriptor $\fy$,  the similarity score between $\fx$ and $\fy$ are defined as the number of matching pair $\fx$ has, \ie,
{\footnotesize
\begin{equation} \label{eq:sim}
\s(\fx,\fy) = \sum_{\vx_i\in \fx, \delta(\vx_i,\fy)<\alpha} 1
\end{equation}
}

Given a querying image,
 matching images with high similarity score can be searched in a database.

%
\CUTTaeho{
For binary feature vectors,
 the distance is measured by hamming distance in the bitwise way,
 $\sdis(\vx_i,\vy_j)=\sum_k \vx_i(k)\oplus \vy_j(k)$,
 here $\vx_i(k)$ is the $k$-th bit of $\vx_i$ and $\oplus$ is the bitwise XOR operator.
For each vector in $\fx$,
 the nearest and second nearest vector in $\fy$ is selected.
Only if the distance from the nearest neighbor is less than a
threshold $\alpha \in(0,1)$ times the distance
 from the second nearest neighbor, a matching pair of vectors is detected.
Then the similarity between descriptors $\fx$ and $\fy$ is the number
 of their matching pairs, \ie,
In this way,
 given a querying image,
 matching images can be searched in a database.
} 

\subsubsection{Cryptographic Tools}

Our system also takes advantage of rich cryptographic algorithms for
privacy protection in cloud-based image search.
It includes: homomorphic encryption, attribute based encryption and
oblivious transfer.

\paragraph{Homomorphic Encryption.}
We employ Paillier's cryptosystem \cite{paillier1999public} as a
building block which has the following
homomorphism\footnote[3]{Computation is conducted in a finite cyclic
  group, and modular operations are followed.
 We omit the modular operations for the sake of  simplicity
  and defer the detailed description on the finite group selection to
  Section \ref{sec:eva}}: $\enh(m_1)\enh(m_2)=\enh(m_1+m_2)$ and $\enh(m_1)^{m_2}=\enh(m_1m_2)$,
where $\enh(m)$ denotes the ciphertext of $m$. Paillier's cryptosystem is proven to be semantically secure against chosen plaintext attack (SS-CPA), which implies that any ciphertext of any message is indistinguishable to a randomly chosen element among the ciphertext space.

\CUTTaeho{
{\small
\begin{table}[h]
\caption{The fast variant of Paillier's Cryptosystem}\label{table:paillier}
\centering
\begin{tabular}[width=\textwidth]{|l|}
\hline
Choose two prime numbers $p$ and $q$.\\
\textbf{Public key}: \\
modulus $n=pq$ and base $g \in \mathbb{Z}_{n^2}^*$\\
\textbf{Private key}: \\
$\lambda=LCM(p-1,q-1)$\\
\textbf{Encryption}: \\
$c=\enh(m)=g^{m+nr} \mod  n^2$, where $r$ is a randomizer\\
\textbf{Decryption}: \\
$m=D(c)=\frac{L(c^\lambda \mod  n^2)}{L(g^\lambda \mod  n^2)} \mod n$, $L(x)=\frac{x-1}{n}$\\
\textbf{Homomorphic}:\\
$\enh(m_1)\enh(m_2)=\enh(m_1+m_2)\mod  n^2$\\
$\enh(m_1)^{m_2}=\enh(m_1m_2)\mod n^2$\\
\hline
\end{tabular}
\vspace{-0.15in}
\end{table}
}
}

Note that the numeric type of feature vectors may be real number, but
the Paillier's cryptosystem is based on large integers, therefore we
need to use integers to represent real numbers
first. \ourprotocolNSP~uses the fixed point representation
to represent real numbers rather than
floating-point representation due to its efficiency.


\CUTTaeho{
Given a real number $a$ and a fixed precision $p$, an $m+1$-dimension
binary array $A$ satisfies the following in the fixed point
representation with Two's Complement:

{\footnotesize
\begin{displaymath}
\begin{cases}
A[0]=0\wedge \sum_{k=1}^mA[k]\cdot 2^{-k+p}\approx a  & a\geq 0 \\
 A[0]=1 \wedge -\sum_{k=1}^{m-1}\left(A\left[k\right]\oplus 1\right)\cdot 2^{-k+p}-A[m]\cdot 2^{-k+m}\approx a& a<0\\
\end{cases}
\end{displaymath}
}
The minimum unit of this representation is $2^{p-m}$ (\ie, precision), and the range of this representation is $(-2^{p-1},2^{p-1})$. Then, we use the following integer to represent the real number $a$ (with some errors less than the minimum unit):

\begin{displaymath}
\fpr(a)=\begin{cases}
 \sum_{k=1}^m A[k]\cdot 2^{m-k} & \text{if }A[0]=0 \\
 -\sum_{k=1}^m \left(A\left[k\right]\oplus 1\right)2^{m-k} & \text{if }A[0]=1 \\
\end{cases}
\end{displaymath}

\noindent which is the definition of signed integer with Two's complement.
Note that the addition/subtraction ($x\pm y$), multiplication ($x\cdot y$) and the division ($\frac{x}{y}$) are all elementary arithmetic operations closed in integer domain (\ie, $a/b$ is the quotient of $\frac{a}{b}$). We assume $m$, $p$ are pre-defined parameters based on the range and precision requirements of the application.
} 


\CUTTaeho{ 
In the cloud-based privacy-preserving image search,
 we design our system to outsource as much computation as possible
 to the powerful cloud with as few interactions as possible with the requester.
The main challenges come from the computation cost and non-interactive requirement.
The core computation of the vision-based image search
 is the Euclidean distances among feature vectors.

We design protocols for both real number descriptor based image search
 and binary descriptor based image search.
And our protocol can also solve the visual words based image search.
} 

\paragraph{Ciphertext-Policy Attribute Based Encryption}
We also adopt ciphertext-policy attributed based encryption (CP-ABE)
\cite{bethencourt2007ciphertext} for access control due to its
generality and security. 
Other attribute based encryption methods can also be adopted, \eg, \cite{chase2009improving}.
In the CP-ABE, a trusted authority (not the image service provider)
takes response of generating public parameters.
Given the public parameters, a data owner can encrypt a
message such that only the users satisfying a certain access rule can
decrypt it. Secret keys of users contain attribute values for the key
holders, and the access rule is expressed with boolean
operators (AND, OR \etc) and attribute values. CP-ABE is proven to be
IND-CCA1 secure, which implies the semantic security against chosen
plaintext attack.

\paragraph{Oblivious Transfer}
The $k$-$n$ oblivious transfer (OT) \cite{camenisch2007simulatable} let a receiver obtain any subset of $k$ items from the sender's $n$ items, while the sender remains oblivious of the receiver's selection, and the receiver remains oblivious of other items as well.


\label{sec:design}
\vspace{-0.05in}
\subsection{System Join}
\vspace{-0.05in}
Whenever a new user joins the system, he generates a pair of Paillier Keys $PK,SK$ and picks a random vector $\vr$, which has the same dimension of the feature vector.
Then, he uses CP-ABE to encrypt $PK,SK,\vr$ under the access rule he wishes to enforce (\ie who can search on his images).
He uploads the following to the sharing cloud, which are the \textit{search keys} to be used in the photo searching later.
\begin{displaymath}\footnotesize
\abe\left(\left\lbrace PK,SK,\vr \mod n\right)\right\rbrace
\end{displaymath}

\vspace{-0.1in}
\subsection{Public \& Private Bag Generation}
\vspace{-0.05in}
When an owner wants to upload his photo $I_x$, the ROP $\rop(I_x)$ is selected either automatically or manually, and the image descriptor $\fx$ of ROP is extracted. $\fx$ is a set of fixed-dimension feature vectors $\fx=\left\lbrace\vx_1,\vx_2,\vx_3,\cdots \right\rbrace$. A photo may have several ROPs (several persons in the same photo), but \textit{w.l.o.g} we consider only one ROP per image since multiple ROP is a simple extension. Then, the owner separates the ROP as public ROP $\public\left(\rop\left(I_x\right)\right)$ and secret ROP $\secret\left(\rop\left(I_x\right)\right)$ as in Section~\ref{sec:privacy_storage}, and the following \textit{public bag} is uploaded to the sharing cloud:
\begin{displaymath}
I_{x,\texttt{pub}}=\left\lbrace I_x-\secret\left(\rop\left(I_x\right)\right) \right\rbrace~~\text{(pixel-wise)}
\end{displaymath}
After the public bag is uploaded, the owner encrypts the private part of ROP as the \textit{private bag} using symmetric encryption such as AES-256. Also, for the cloud-based search, he homomorphically encrypts the feature descriptor, which are stored in the \textit{search bag} (Protocol~\ref{private_bag:real}).

\floatname{algorithm}{Protocol}
\begin{algorithm}[t!]\footnotesize
\caption{Secret \& Search bag generation}\label{private_bag:real}
\begin{algorithmic}[1]

\STATE The owner of $I_x$ randomly picks a symmetric key $K_{e}$ and uses symmetric encryption (AES in this paper) to encrypt the private part of the ROP as $\en_{K_e}(\secret(\rop(I_x)))$. $K_e$ is encrypted via CP-ABE under his privacy policy as $\abe\left(K_e\right)$.


\STATE For every dimension $\vx(k)$ in every vector $\vx\in\fx$, he computes the following homomorphic ciphertexts using his $PK$ and $\vr$:
\begin{displaymath}
\enh\left(\left(\vx\left(k\right)\right)^2\right),\enh\left(-\vr(k)\cdot\vx\left(k\right)\right)
\end{displaymath}
\end{algorithmic}
\end{algorithm}
Then, the private bag and the search bag of $I_x$ are:
\begin{displaymath}
\begin{split}
I_{x,\texttt{pri}}&=\left\lbrace \begin{array}{c}
\en_{K_e}\left(\secret\left(\rop\left(I_x\right)\right) \right)\\
\abe\left(K_e\right)
\end{array}\right\rbrace \\
I_{x,\texttt{sch}}&=\left\lbrace
\begin{array}{c}
\enh\left(\fx^2\right)\\\enh\left(-\vr\circ\fx\right)
\end{array}
\right\rbrace
\end{split}\end{displaymath}

\noindent where $\fx^2$ and $-\vr\circ\fx$ represent the sets $\left\lbrace \left(\vx_i\left(k\right)\right)^2 \right\rbrace_{\forall i,k}$ and $\left\lbrace -\vr(k)\cdot\vx_i\left(k\right) \right\rbrace_{\forall i,k}$ (Hadamard product between $-\vr$ and each $\vx_i$) respectively. The private/search bag are uploaded to the sharing/search cloud respectively.


\vspace{-0.05in}
\subsection{Cloud-based Image Search}\label{sec:cloud_based_search}
\vspace{-0.05in}

When a querier wants to search an image $I_y$ among a specific owner's images, he extracts corresponding image descriptor $\fy$ and obtains the owner's search keys $\abe(PK,$ $SK,\vr)$ from the server. If he is authorized to search on the owner's images, he will successfully decrypt the search keys and further proceed. Next, he encodes every single dimension of the feature vectors in the querying image as follows:
\begin{displaymath}\footnotesize
\begin{split}
\cipher_1\left(\vy_j\left(k\right)\right)&=\vr(k)^{-1}\cdot \vy_j(k)\\
\cipher_2\left(\vy_j\left(k\right)\right)&=\enh\left(\left(\vy_j\left(k\right)\right)^2\right)
\end{split}
\end{displaymath}

Consequently, the querier achieves two sets of encoded feature descriptors $\cipher_1\left(\fy\right),\cipher_2\left(\fy\right)$ corresponding to $I_y$. He then sends these two sets to the cloud server to outsource the image search. After receiving the encoded descriptors, the cloud conducts several homomorphic operations to achieve the encrypted pairwise distances between $\vx_i$ and $\vy_j$ for all $i,j$ with $I_{x,sch}$ in the search cloud (Protocol~\ref{protocol_closest_proximity:real}). Then, he sends all the ciphertexts of results back to the querier.

\floatname{algorithm}{Protocol}
\begin{algorithm}[t!]\footnotesize
\caption{Privacy-preserving Distance Calculation}
\begin{algorithmic}[1]
\STATE The cloud conducts the following homomorphic operations for all $k$:\vspace{-5pt}
\begin{displaymath}
\begin{split}
&\enh\left(-\vr(k)\cdot\vx_i\left(k\right)\right)^{2\cipher_1\left(\vy_j\left(k\right)\right)}
=\enh\left( -2\vx_i\left(k\right)\vy_j\left(k\right)\right),\\
&\enh\left(\left(\vx_i\left(k\right)\right)^2\right)\cdot \enh\left( -2\vx_i\left(k\right)\vy_j\left(k\right)\right)\cdot \cipher_2\left(\vy_j\left(k\right)\right)\\
=&\enh\left(\left(\vx_i(k)-\vy_j(k)\right)^2\right)
\end{split}
\end{displaymath}

\STATE Then, he computes:\vspace{-8pt}
\begin{displaymath}\scriptsize
\begin{split}
\prod_k{\enh\left(\left(\vx_i(k)-\vy_j(k)\right)^2\right)}==\enh\left(\dis^2\left(\vx_i, \vy_j\right)\right)
\end{split}\end{displaymath}
\end{algorithmic}
\label{protocol_closest_proximity:real}
\vspace{-0.1in}
\end{algorithm}
Upon receiving the ciphertexts of pair-wise distances, the querier uses $SK$ to decrypt every $\dis^2\left(\vx_i,\vy_j\right)$. Then, he finds the top-2 nearest distances to compute the similarity scores between feature descriptor $\fx$ and every $\fy$ according to Eq.~\ref{eq:sim}.

\vspace{-0.05in}
\subsection{Image Retrieval}
\vspace{-0.05in}
Based on the similarity scores, the querier requests the public bags
as well as the private bags of the top-$k$ similar images from the
sharing cloud (\eg, by requesting the URLs). However, explicit request
reveals the search result to the server. Even if every secret part of
ROP is encrypted and the query contents are well protected, cloud may
infer side information by gathering the statistics of the image
retrieval (\eg, popular images and frequently visited images). Thus,
we need to hide the retrieval pattern as well.

To achieve this requirement, we employ the $k$-$n$ OT
(Section~\ref{sec:preliminary}). Since it is extremely expensive to
construct a $k$-$n$ OT with a large $n$, we do not directly run a
$k$-$n$ OT across the whole database to obliviously retrieve $k$
images. Instead, we try to find a trade-off between privacy and
performance as follows. The querier determines a random subset
$\sigma'\subseteq DB$ which contains the set of images $\sigma$ that
he wants to retrieve. The sizes of $\sigma$ and $\sigma'$ are $k$ and
$n$ respectively. Then, the querier and the sharing cloud engage in a
$k$-$n$ OT to let the querier obliviously select the $k$ images.

\CUTTaeho{
 and the construction is usually expensive.
In our system,
 the server has $N$ image records $\mathbf{R}=\{R_1, \cdots R_N\}$.
Each record $R_i$ contains the URL to its public part
 and information of its private part.
The requestor has a secret  subset $\sigma \subseteq \{1,2 \cdots,N\}$
  which are the indices of his desired images computed using image
  matching method, and $|\sigma| = k$.
To achieve the requestor's $k$-out-of-$n$ selection oblivious to the server,
 the requestor can use $\sigma'$ with randomly added indices to blind the server,
 here $\sigma \subset \sigma'$ and $|\sigma'|=n$.
Note that $n$ is not equal $N$,
 instead, for large dataset with big $N \gg n$.

With the encoded request and the oblivious image retrieval strategy,
 \ourprotocol hides the search content from untrusted parties.
}

\CUTTaeho{


%
%


\vspace{-0.05in}
\subsubsection{Privacy Preserving Image Search}
\vspace{-0.05in}
When a requester needs to search an image of a specific owner,
 he obtains the owner's $\abe\left(\left\lbrace PK,SK,K^0,K^1,s\right\rbrace\right)$ from the server
 and tries to decrypt them with his attributes. A requester with valid attribute  successfully decrypts correct $PK,SK,K^0,K^1$ and $s$.
Then, for each bit $\fy(k)$ of his request feature descriptor $\fy$, he uses $H^k(s)K^0$ and $H^k(s)K^1$ to encode the bit value as the garbled input $\code(\fy(k))$. Then, he achieves:
\begin{displaymath}\footnotesize
\code(\fy) = \left\lbrace\code\left(\vy_1\right), \cdots, \code\left(\vy_n\right)\right\rbrace
\end{displaymath}
where each garbled vector is:
\begin{displaymath}\footnotesize
\code(\vy_1)=\left(H\left(s\right)K^{\vy_1\left(1\right)},H^2\left(s\right)K^{\vy_1\left(2\right)},
H^3\left(s\right)K^{\vy_1\left(3\right)},\cdots\right)
\end{displaymath}
Then, the requester uploads $\code(\fy)$ to the cloud server.

The cloud server, upon receiving any garbled vector $\code(\vy_j)$, can efficiently achieve $\enh\left(\sdis\left(\vx_i,\vy_j\right)\right)$ without interacting with the requester or the image owner (Protocol \ref{protocol_closest_proximity}).

The cloud server runs Protocol~\ref{protocol_closest_proximity} to calculate the encrypted pairwise distances between $\vx_i$ and $\vy_j$ for all $i,j$ and sends the list back to the requester.

\floatname{algorithm}{Protocol}
\begin{algorithm}[h]\footnotesize
\caption{Privacy-preserving Distance Calculation}
\textbf{Input}: $\garbled\left(\vx_i\left(k\right)\right)$, $\code\left(\vy_j\left(k\right)\right)$ for all $k$\\
\textbf{Output}: $\enh\left(\sdis\left(\vx_i, \vy_j\right)\right)$
\begin{algorithmic}[1]
\STATE For every garbled gate $\garbled\left(\vx_i\left(k\right)\right)$, the cloud server looks up and symmetrically decrypts $\enh\left(\vx_i\left(k\right)\oplus\vy_j\left(k\right)\right)$ from the shuffled table.

\STATE Then, he computes:
\begin{displaymath}\vspace{-4pt}
\begin{split}
\prod_k{\enh\left(\vx_i\left(k\right)\oplus\vy_j\left(k\right)\right)}
&=\enh\left(\sum_k{\vx_i\left(k\right)\oplus\vy_j\left(k\right)}\right)\\
&=\enh\left(\sdis\left(\vx_i,\vy_j\right)\right)
\end{split}\end{displaymath}

\end{algorithmic}
\label{protocol_closest_proximity}
\end{algorithm}

Then, the requester decrypts the distances using $SK$, and for each image $I_x$, he conducts a simple top-2 computing and threshold comparison
 to learn the similarity between feature descriptor $\fx$ and $\fy$
 according to Eq.~\ref{eq:sim}.
The requester decides which images to retrieve according to their similarity to the request image.
}


\vspace{-0.08in}
\section{Security Analysis and Refinement}

\label{sec:discussion}

\vspace{-0.05in}
\subsection{Security Analysis}
\vspace{-0.05in}

Firstly, the secret part of ROP is well protected by the symmetric encryption, whose key is encrypted with CP-ABE proven to be semantically secure. Besides, the search keys are also protected by the CP-ABE. Therefore, clouds cannot infer sensitive information from its storage in \ourprotocolNSP.

Then, we design a game to prove that \ourprotocolNSP~ reveals no sensitive information to the cloud servers during the photo searching procedure theoretically.
We omit the proof here due to space limitation,
 readers can refer to the appendix of (\cite{appendix}) for the detail of proof.

\CUTTaeho{
\begin{proof}
We define two PPTAs $\adversary_1,\adversary_2$, and define their advantages $\textsf{adv}_{i}$ as:
\begin{displaymath}
\textsf{adv}_{i}=\text{Pr}\left[
\begin{array}{c}
\adversary_i\leftarrow\cipher_i\left(\fpr\left(\fy\right)\right)\\
y_i'=y
\end{array}
\right]-\frac{1}{2}
\end{displaymath}

\noindent That is, $\textsf{adv}_i$ is the advantage of $\adversary_i$ when he is only given $\cipher_i\left(\fpr\left(\fy\right)\right)$ and gives a guess $y_i'$ on $y$. Since $\adversary$ is given both adversaries' views, if $\adversary_1,\adversary_2$ agree on the same guess, he will also give the same guess, otherwise his advantage does not change. Then, we have the following probabilities for four cases:
\begin{displaymath}
\begin{split}
\text{Pr}\left[y'=y|y_1'=y\wedge y_2'=y\right]&=1\\
\text{Pr}\left[y'=y|y_1'=y\wedge y_2'\neq y\right]&=\text{Pr}\left[y'=y\right]\\
\text{Pr}\left[y'=y|y_1'\neq y\wedge y_2'=y\right]&=\text{Pr}\left[y'=y\right]\\
\text{Pr}\left[y'=y|y_1'\neq y\wedge y_2'\neq y\right]&=0
\end{split}
\end{displaymath}

Since $\adversary_1$ and $\adversary_2$ gives their guesses based on independent views, we have
\begin{displaymath}
\begin{split}
\text{Pr}\left[y_1'=y\wedge y_2'=y\right]&=\text{Pr}\left[y_1'=y\right]\cdot \text{Pr}\left[y_2'=y\right]\\
\text{Pr}\left[y_1'\neq y\wedge y_2'=y\right]&=\text{Pr}\left[y_1'\neq y\right]\cdot \text{Pr}\left[y_2'=y\right]\\
\text{Pr}\left[y_1'=y\wedge y_2'\neq y\right]&=\text{Pr}\left[y_1'=y\right]\cdot \text{Pr}\left[y_2'\neq y\right]\\
\text{Pr}\left[y_1'\neq y\wedge y_2'\neq y\right]&=\text{Pr}\left[y_1'\neq y\right]\cdot \text{Pr}\left[y_2'\neq y\right]\\
\end{split}
\end{displaymath}
Note that
\begin{displaymath}
\textsf{adv}=\text{Pr}\left[y'=y\right]-\frac{1}{2}=1-\text{Pr}\left[y'\neq y\right]-\frac{1}{2}
\end{displaymath}
Then, given those conditional probabilities, the total probability is:

\CUTTaeho{
According to the definition of the advantages,
\begin{displaymath}
\textsf{adv}=\text{Pr}\left[y'=y\right]-\frac{1}{2}=1-\text{Pr}\left[y'\neq y\right]-\frac{1}{2}
\end{displaymath}
}

\begin{displaymath}
\begin{split}
\text{Pr}\left[y'=y\right]&=\frac{1}{2}+\textsf{adv}\\
&=1\cdot \left(\frac{1}{2}+\textsf{adv}_1\right)\left(\frac{1}{2}+\textsf{adv}_2\right)\\
&+\left(\frac{1}{2}+\textsf{adv}\right)\left(\frac{1}{2}+\textsf{adv}_1\right)\left(\frac{1}{2}-\textsf{adv}_2\right)\\
&+\left(\frac{1}{2}+\textsf{adv}\right)\left(\frac{1}{2}-\textsf{adv}_1\right)\left(\frac{1}{2}+\textsf{adv}_2\right)\\
&+0
\end{split}
\end{displaymath}

\noindent which leads to
\begin{displaymath}
\textsf{adv}=\frac{\left(\frac{1}{2}+\textsf{adv}_1\right)\left(\frac{1}{2}+\textsf{adv}_2\right)}{\frac{1}{2}-2\textsf{adv}_1\textsf{adv}_2}-\frac{1}{2}
\end{displaymath}

Both Paillier's cryptosystem and CP-ABE are proved to be semantically secure against chosen plaintext attack (SS-CPA)\footnote[3]{CP-ABE is proved to achieve IND-CPA, which implies SS-CPA} \cite{paillier1999public,bethencourt2007ciphertext}. Therefore, $\adversary$ does not have a significant chance to get $SK,r^{-1}$ in $\abe\left(PK,SK,r^{-1}\right)$ or $\fpr\left(\fy\right)$ in $\cipher_2\left(\fpr\left(\fy\right)\right)$, which means $\textsf{adv}_2$ is negligible. Recall that the function family $x\rightarrow \mu x\mod n$ is $\epsilon$-pairwise independent for negligible $\epsilon$, and $\mu^{-1}x\mod n$ is close to uniform in $\mathbb{Z}_n$. Therefore, he does not have a significant chance to get $\fpr\left(\vy_j\left(k\right)\right)$ in $\cipher_1\left(\fpr\left(\vy_j\left(k\right)\right)\right)$ either, which implies a negligible $\textsf{adv}_1$. Since both $\textsf{adv}_1,\textsf{adv}_2$ are negligible, $\textsf{adv}$ is negligible too.
\end{proof}\medskip

}

Besides the adversarial cloud servers, we have also assumed malicious
queriers in our adversarial model. However, unauthorized malicious users
are not as threatening as cloud servers since they never get involved
in any transaction with valid users. All they can do except
compromising the server is to try man-in-the-middle attacks to sniff
the search results, but this can be trivially prevented by introducing
secure communication channel. Even if they compromised a server,
CP-ABE guarantees the indistinguishability of the ciphertexts. In
conclusion, malicious users do not learn about sensitive information
either.

\vspace{-0.05in}
\subsection{Refinements for Binary Descriptor}
\vspace{-0.05in}
Some image retrieval systems use binary image feature descriptors
 because they are more compact and computationally manageable than real number ones,
 with a little accuracy loss in content recognition \cite{alahi2012freak}.
It is more suitable for resource-limited applications. However,
directly applying \ourprotocolNSP~in mobile platforms with binary
descriptors does not fully exploit the advantage of it. The
exponentiation operations contribute to majority of the computation
overhead in our cryptographic building blocks, but both image owners
and queriers need $\Theta(\alpha)$ exponentiations
throughout the protocol where $\alpha$ is the number of interest
points in a image.

To relax this bottleneck, we further design our framework for the special case where binary descriptors are used (refer to the system as $\ourprotocol_{bin}$),
Note that for any two vectors $\vx,\vy$, we have:
\begin{displaymath}\footnotesize
\dis^2\left(\vx,\vy\right)=\sum_k\left(\vx\left(k\right)-\vy\left(k\right)\right)^2=\sum_k\vx\left(k\right)\oplus\vy\left(k\right)
\end{displaymath}
\noindent where $\vx(k)$ is the $k$-th bit of $\vx$ and $\oplus$ is the bitwise XOR operator. Therefore, we consider using a succinct garbled circuit in combination with homomorphic encryption to achieve a light-weight and non-interactive framework dedicated to binary descriptor based search,
which is one of our contributions.


\subsubsection{Yao's Garbled Circuit}

To enhance the understanding, we briefly review Yao's garbled circuit (GC), and we direct the readers to relevant literal works \cite{lindell2004proof} for technical details. Yao's Garbled Circuit is designed for two-party computation,
 where $P_x$ and $P_y$ wish to jointly compute a function $F$
 over their private input $\vx$ and $\vy$ using a garbled boolean circuit.
Here we use an XOR gate as an example.
\vspace{3pt}

{\centering\footnotesize
\begin{minipage}{0.13\textwidth}
  \centering
\includegraphics[width=0.5\linewidth, clip,keepaspectratio]{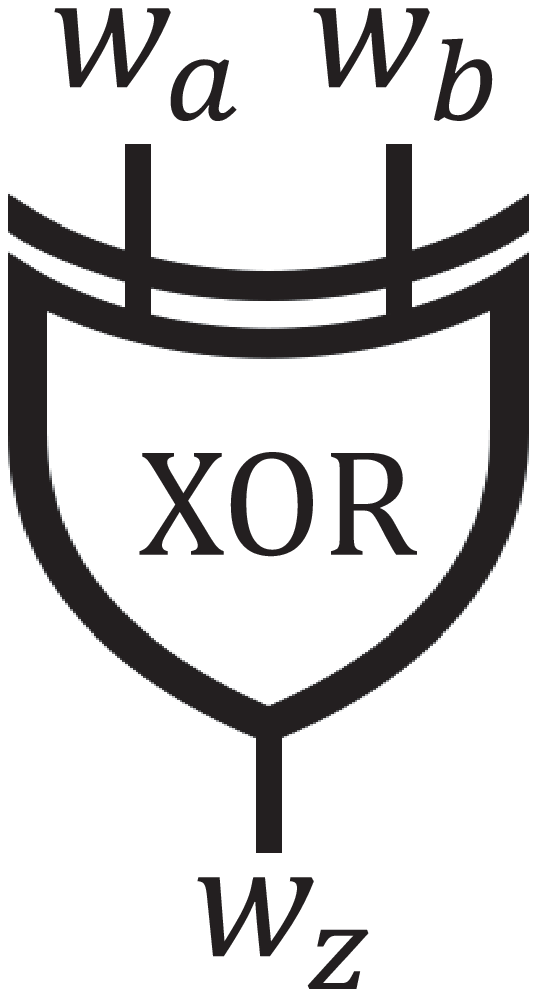}
\label{fig:circuit}
\vspace{-7pt}\captionof{figure}{Gate}
\end{minipage}
\begin{minipage}{0.36\textwidth}
  \centering
\begin{tabular}{|c|c|c|}
\hline
$k_a^0$ & $k_b^0$ & $\en_{k_a^0}(\en_{k_b^0}(k_z^0))$\\\hline
$k_a^0$ & $k_b^1$ & $\en_{k_a^0}(\en_{k_b^1}(k_z^1))$\\\hline
$k_a^1$ & $k_b^0$ & $\en_{k_a^1}(\en_{k_b^0}(k_z^1))$\\\hline
$k_a^1$ & $k_b^1$ & $\en_{k_a^1}(\en_{k_b^1}(k_z^0))$\\\hline
\end{tabular}
\captionof{table}{Garbled Gate $\garbled(w_a\oplus w_b)$}\label{table:GC}
\end{minipage}
}

Two random values $k_i^0$,$k_i^1$ are chosen to represent the bit
values 0 and 1 for each wire $w_i$. Then, the shuffled
Table~\ref{table:GC} represents the garbled XOR gate (shuffled so that
inputs are not inferred from the row number). Given two garbled
inputs, the evaluator can obliviously evaluate the boolean gate by
looking up the shuffled table and decrypting the output to get a
garbled output.


\subsubsection{System Join}

The new joiner generates a pair of Paillier keys $PK,SK$ and picks two symmetric encryption keys $K^0,K^1$ as well as a random seed $s$. Then, he defines a privacy policy to specify which group of people are authorized to search on his images. $PK,SK,K^0,K^1$ and $s$ are encrypted using CP-ABE as $\abe\left(\left\lbrace PK,SK,K^0,K^1,s \right\rbrace\right)$, which are uploaded to the sharing cloud as his search keys. Finally, he uses $PK$ to encrypt 0,1 homomorphically for later use, \ie, $\enh(0),\enh(1)$.

\subsubsection{Public \& Private Bag Generation}

To upload a photo $I_x$, the owner extracts the ROP $\rop\left(I_x\right)$ as well as the binary image descriptor $\fx,$ and generates the public bag as in the original framework \ourprotocolNSP. After uploading the public bag to the sharing cloud, he symmetrically encrypts the private part of ROP, and keeps it as well as the key in the private bag. Then, he uses a collision-resistant hash function $H(\cdot)$ and the search keys to garble each bit as a garbled gate (Protocol~\ref{private_bag:binary}), where $H^k(\cdot)$ denotes applying the hash function for $k$ times.

\floatname{algorithm}{Protocol}
\begin{algorithm}[t!]\footnotesize
\caption{Secret \& Search Bag Generation}\label{private_bag:binary}
\begin{algorithmic}[1]

\STATE The owner randomly picks a key $K_{e}$ and uses symmetric encryption (AES in this paper) to encrypt the private part of the ROP as $\en_{K_e}(\secret(\rop(I_x)))$. $K_e$ is encrypted via CP-ABE under his privacy policy as $\abe\left(K_e\right)$.

\STATE For every bit $\vx(k)$ in every vector $\vx\in\fx$, he generates and shuffles the following table:

{\footnotesize
\begin{center}
\begin{tabular}{|c|c|}
\multicolumn{2}{c}{if $\vx(k)=0$}\\\hline
$\gamma_0 = H^k(s)\cdot K^0$ & $\en_{\gamma_0}\left(\enh\left(0\right)\right)$ \\
$\gamma_1 = H^k(s)\cdot K^1$ & $\en_{\gamma_1}\left(\enh\left(1\right)\right)$\\\hline
\multicolumn{2}{c}{if $\vx(k)=1$}\\\hline
$\gamma_0 = H^k(s)\cdot K^0$ & $\en_{\gamma_0}\left(\enh\left(1\right)\right)$ \\
$\gamma_1 = H^k(s)\cdot K^1$ & $\en_{\gamma_1}\left(\enh\left(0\right)\right)$\\\hline
\end{tabular}
\end{center}
}

\noindent which represents $\vx(k)$'s garbled gate $\garbled\left(\vx\left(k\right)\right)$.
\end{algorithmic}
\end{algorithm}

\begin{figure}[h]\vspace{-7pt}
\begin{center}
\includegraphics[width=0.8\linewidth]{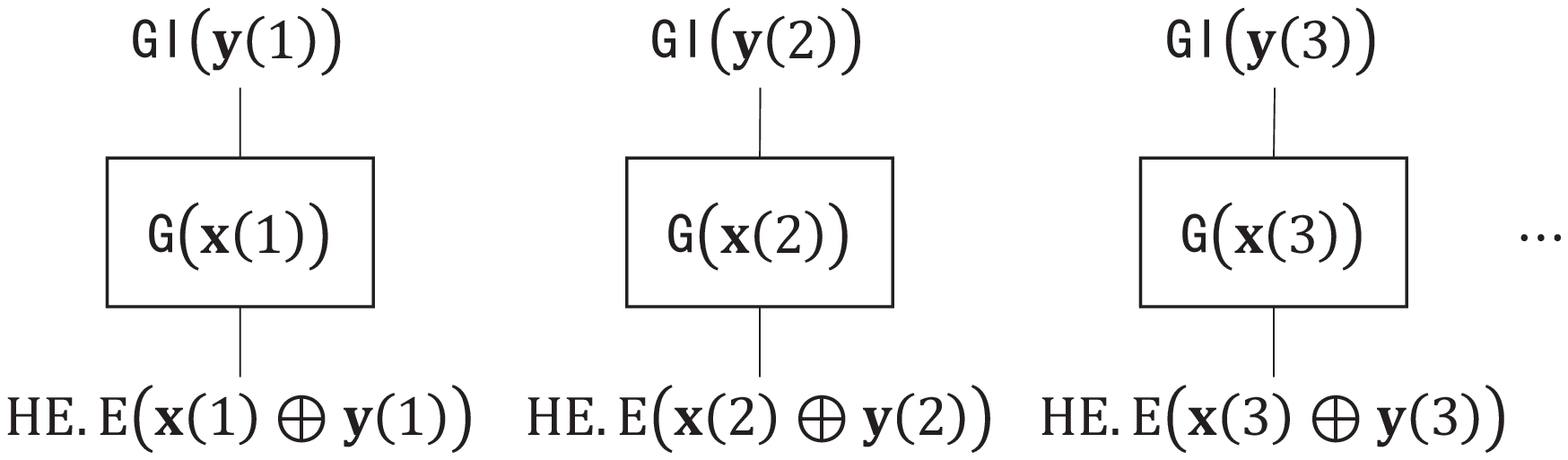}
\end{center}
\vspace{-10pt}\caption{Garbled gates $\garbled(\vx)$ from Protocol~\ref{private_bag:binary}}\vspace{-10pt}
\label{fig:gate}
\end{figure}

From the protocol, the feature vector $\vx$ is encrypted to a series of garbled gates (Fig.~\ref{fig:gate}),
and the following are corresponding private bag and search bag of $I_x$:
\begin{displaymath}
\begin{split}
I_{x,\texttt{pri}}&=\left\lbrace \begin{array}{c}
\en_{K_e}\left(\secret\left(\rop\left(I_x\right)\right) \right)\\
\abe\left(K_e\right)
\end{array}\right\rbrace \\
I_{x,\texttt{sch}}&=\left\lbrace
\garbled\left(\fx\right)
\right\rbrace
\end{split}\end{displaymath}

\subsubsection{Cloud-based Image Search}

To search a photo $I_y$ from other's ones, the querier extracts corresponding descriptor $\fy$ and obtains the owner's $\abe\left(\left\lbrace PK,SK,K^0,K^1,s\right\rbrace\right)$. If he successfully decrypts it, he further uses $H^k(s)K^0$ or $H^k(s)K^0$ to encode each $k$-th bit $\vy(k)$ as the garbled input $\code\left(\vy\left(k\right)\right)$ to finally achieve the set of garbled inputs $\code\left(\fy\right)$, which is uploaded to the cloud.
\CUTTaeho{
\begin{displaymath}
\code(\fy) = \left\lbrace\code\left(\vy_1\right), \cdots, \code\left(\vy_n\right)\right\rbrace
\end{displaymath}
where each garbled vector is:
\begin{displaymath}
\code(\vy_1)=\left(H\left(s\right)K^{\vy_1\left(1\right)},H^2\left(s\right)K^{\vy_1\left(2\right)},
H^3\left(s\right)K^{\vy_1\left(3\right)},\cdots\right)
\end{displaymath}
}
The cloud server conducts homomorphic operations to achieve $\enh\left(\dis^2\left(\vx_i,\vy_j\right)\right)$ for all $i,j$ without interacting with the requester or the image owner (Protocol \ref{protocol_closest_proximity}). Then, he sends the ciphertexts back to the querier, who proceeds as \ourprotocolNSP.
\floatname{algorithm}{Protocol}
\begin{algorithm}[h]\footnotesize
\caption{Privacy-preserving Distance Calculation}
\begin{algorithmic}[1]
\STATE For every garbled gate $\garbled\left(\vx_i\left(k\right)\right)\in\garbled\left(\vx_i\right)$, the cloud server looks up and symmetrically decrypts $\enh\left(\vx_i\left(k\right)\oplus\vy_j\left(k\right)\right)$ from the shuffled table.

\STATE Then, he computes:\vspace{-4pt}
\begin{displaymath}\footnotesize
\begin{split}
\prod_k{\enh\left(\vx_i\left(k\right)\oplus\vy_j\left(k\right)\right)}
&=\enh\left(\sum_k{\vx_i\left(k\right)\oplus\vy_j\left(k\right)}\right)\\
&=\enh\left(\dis^2\left(\vx_i,\vy_j\right)\right)
\end{split}\end{displaymath}

\end{algorithmic}
\label{protocol_closest_proximity}
\end{algorithm}

\CUTTaeho{
\subsection{Linear Search and Optimization}
\label{sec:optimize}

When computing the matching score, our protocol simply conducts a
linear search on every feature vector in an image descriptor. In fact,
this may lead to a performance bottleneck when the number of feature
vectors is large. However, the basic component of existing image
search techniques is two vectors' Euclidean distance calculation, and
there exists various optimized methods which are based on vector
distance calculation. Although our protocol is based on the linear
search, one can easily improve our solution with existing
optimizations to achieve better performance. We briefly review
possible options here.

\textbf{K-D Tree}:
For some complicated images with hundreds of feature vectors,
 K-dimensional tree (k-d tree) based fast approximate nearest neighbors algorithms \cite{muja2009fast} are more useful than linear search,
 are extensively used in the computer vision field. K-d tree can speed up the matching of high-dimensional vectors
 by up to several orders of magnitude compared to linear search
 with little accuracy loss.

\CUTTaeho{
The original k-d tree based method splits the data in half at each level of the tree on the dimension
for which the data exhibits the greatest variance.
When searching the trees, a single priority queue
is maintained across all the randomized trees so that
search can be ordered by increasing distance to each
bin boundary.
}

\textbf{Visual Words}:
Visual words are recently proposed for large-scale image search.
The space of descriptors is quantized
 to obtain the visual vocabulary (this process is performed off-line),
 which reduces the cardinality of the vector space. Then, value of each dimension is drawn from the vocabulary, and an image is represented by the
frequency histogram of visual words obtained by assigning
each descriptor of the image to the closest visual word.

}

\CUTTaeho{
\subsubsection{Distributed Computation}
Distributed computing such as MapReduce can be utilized  is a software framework
 for the applications which process vast amounts of data
 on large clusters of commodity hardware in a parallel manner.
Exploiting the power of parallel computing cloud,
 the image query can be handled within an acceptable response time.
}


\vspace{-0.08in}
\section{Implementation and Evaluation}

\label{sec:eva}

\CUTTaeho{
\begin{figure}[t!]
\centering
\includegraphics[width=0.9\linewidth, clip,keepaspectratio]{public.eps}
\caption{Thumbnails of sample public and private images generated by three concealing
methods.}
\label{fig:public}
\end{figure}
}
\vspace{-0.05in}
\subsection{Development Environment}\label{sec:evaluation}
\vspace{-0.05in}
We implemented both client side and cloud side of \ourprotocolNSP.
The client side program is developed for Android smartphones and the commodity laptops for performance
comparisons, and the cloud side program is developed only for the laptops. We used HTC G17 (1228Hz CPU, 1G
RAM) and ThinkPad X1 (i7, 2.7GHz CPU, 4G RAM).

The CP-ABE is implemented based on the PBC library, and other building blocks (Section\ref{sec:preliminary}) are implemented in Java,
 including the AES (128-bit), Paillier's cryptosystem (512-bit primes $p,q$), $k$-$n$ oblivious transfer and the fixed point operations.
 Based on these building blocks, we implemented the core protocols in both variants $\ourprotocolNSP$ and $\ourprotocolNSP_{bin}$.
The automatic ROP detection is implemented with cascade object detection (\eg, face detection) \cite{viola2004robust}.
We employed widely used 64-dimensional SURF descriptor \cite{bay2008speeded} and 128-dimensional SIFT descriptor \cite{lowe2004distinctive} for
the variant of real number descriptors (\ourprotocolNSP~), and 64 bit binary SURF and 128 bit binary SIFT for $\ourprotocolNSP_{bin}$.
Although our evaluation is conducted with these descriptors,
 our system is compatible with other vector-based descriptors too.
Both the object detection and descriptor extraction
 are implemented using the image process library OpenCV 
 for Window and Android.
ROP separation (\textbf{Mask}, \textbf{P3\cite{ra2013p3}}, and \textbf{Blur}) is also implemented with it.


\vspace{-0.05in}
\subsection{Real-life Datasets}
\vspace{-0.05in}
To measure the privacy protection and the cost of \ourprotocolNSP,
 we used the well-known Labelled Faces in the Wild (LFW) dataset \cite{LFWTech},
 which consists of 30,281 real-life images collected from news photographs.
We detect all human faces automatically and set those faces as ROPs of images, and $9$ feature vectors are extracted as their image descriptor \cite{luo2007person}. On average,
ROP occupies less than 20\% of each image for 80\% images.
We also used the INRIA Holidays dataset \cite{jegou2008hamming},
 which contains 1,491 high-resolution personal photos taken during their holidays (majority with resolution 2560px$\times$1920px).
We set the entire image of the INRIA as the ROP.







\CUTXY
{
\begin{figure}[t]
\begin{center}
\begin{minipage}[t]{0.45\linewidth}
\centering
\includegraphics[width=\linewidth, clip,keepaspectratio]{featurepoint.eps}
\caption{Privacy on Feature Detection Algorithm.}
\label{fig:feature}
\end{minipage}
\quad
\begin{minipage}[t]{0.45\linewidth}
\centering
\includegraphics[width=\linewidth, clip,keepaspectratio]{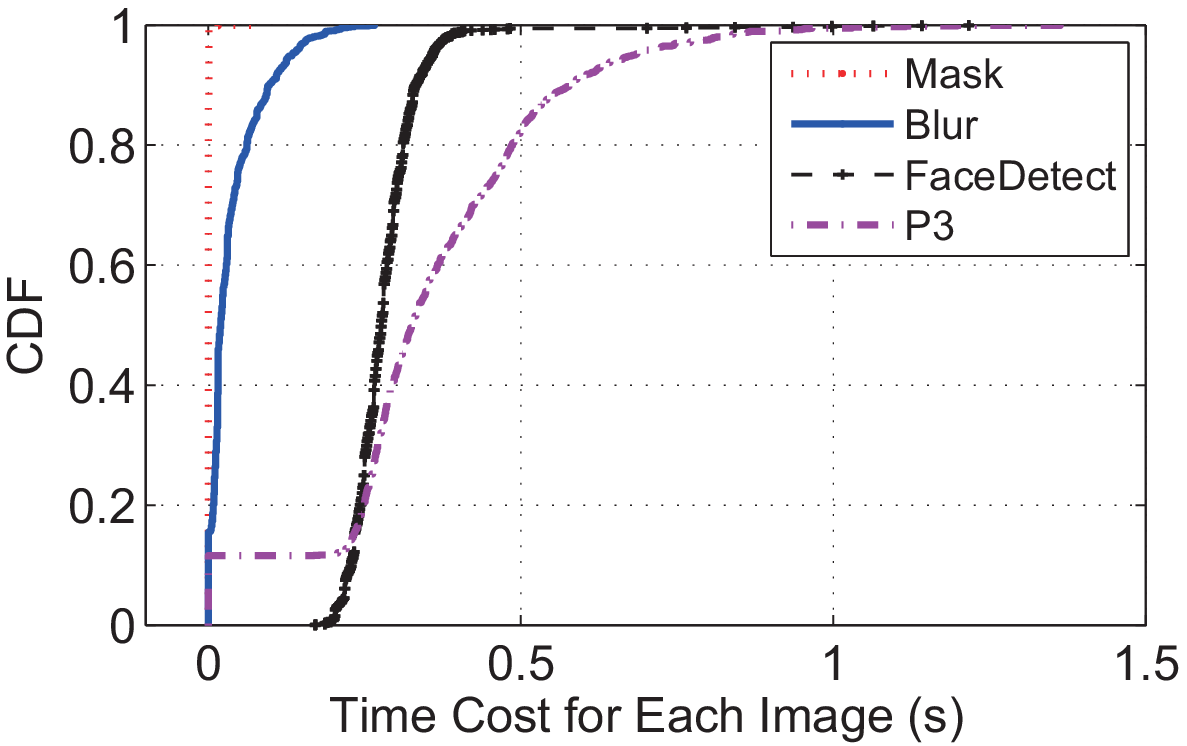}
\caption{Time cost of public part generation for each
  image.}
\label{fig:concealtime}
\end{minipage}
\end{center}
\end{figure}
}




\subsection{Image Recognition on Public Part}

ROPs are separated in three different methods (Mask, P3 and Blur) respectively.
 To evaluate the safety against the object detection algorithms,
 we ran face detection \cite{viola2004robust} and feature points detection \cite{lindeberg1998feature} algorithms on the public part of ROPs.
On average,
 there are $1.1$ faces in each original image in LFW, but only $0.017$, $0.029$ and $0.028$ faces are detected in the public part of the ROPs generated by Mask, P3 and Blur respectively, and our manual examination shows that majority of the detections were false positives (\eg, some textures being detected as faces). Therefore, we conclude that almost \textbf{no faces} are detected in the public parts of images by algorithm.
 Also, \textbf{no matched feature points} are detected in the public parts of ROPs
 for both LFW and Holiday datasets as well.
As a conclusion, all three methods provide good privacy protection against face/feature detection algorithms.

\begin{figure}[h]
 \centering
  \subfigure[LFW]{
    \label{fig:concealtimeLFW}
    \includegraphics[width=0.42\linewidth, clip,keepaspectratio]{timecost.eps}}
  \subfigure[Holiday]{
    \label{fig:concealtimeH}
   \includegraphics[width=0.42\linewidth, clip,keepaspectratio]{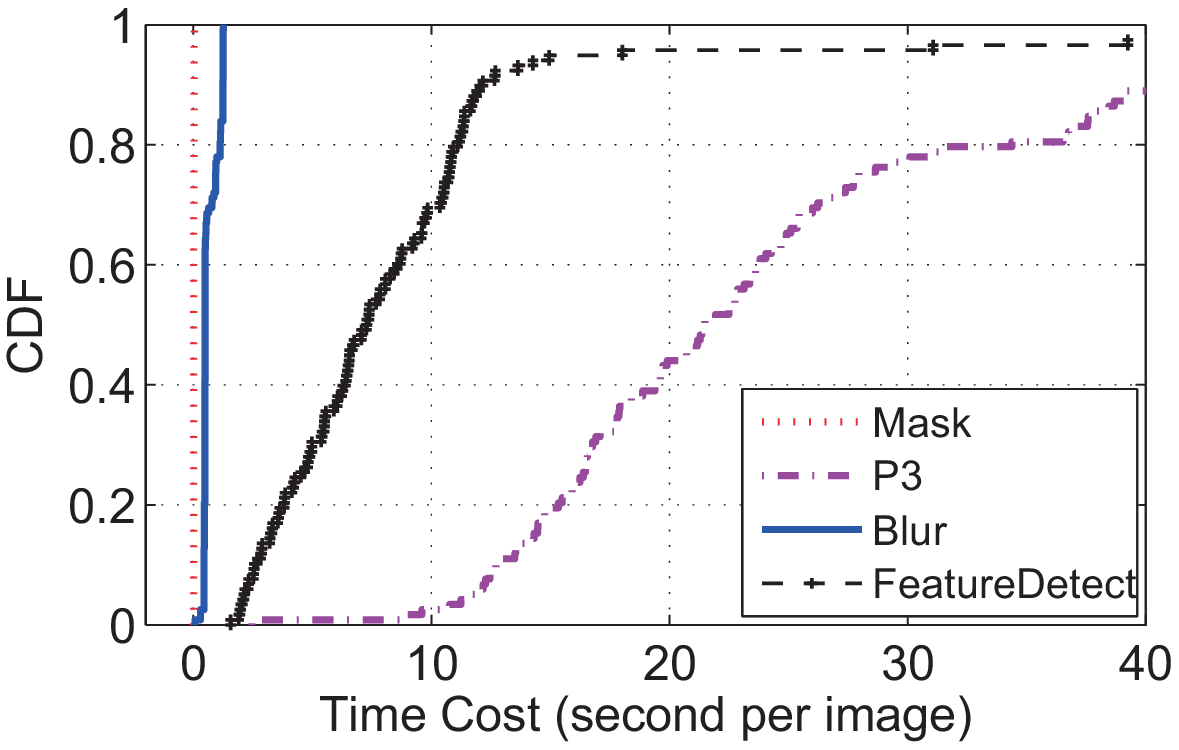}}
      \vspace{-5pt}\caption{Run time of ROP separation}\vspace{-5pt}
  \label{fig:concealtime}
\end{figure}

\begin{figure*}[t]
  \centering
  \subfigure[Public]{
    \label{fig:storage:pub}
    \includegraphics[width=0.2\linewidth, clip,height=0.8in]{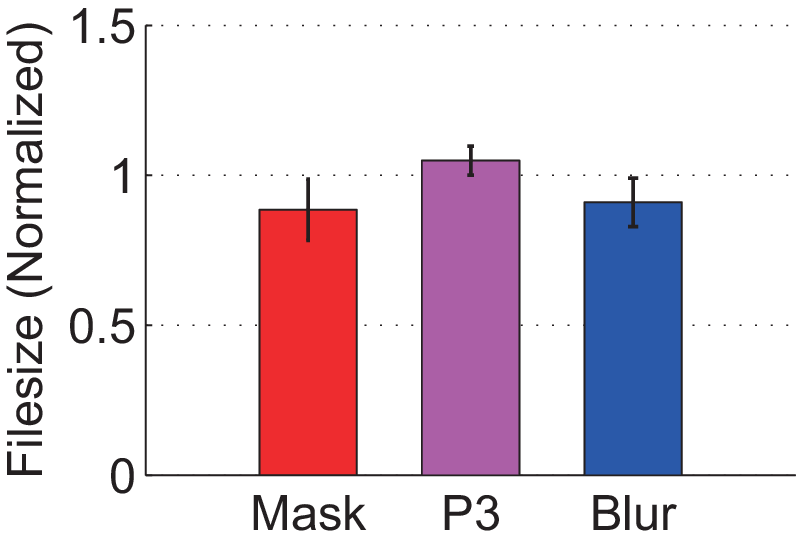}}
    \ \ \ \ \ \ \ \
  \subfigure[Secret]{
    \label{fig:storage:pri}
   \includegraphics[width=0.2\linewidth, clip,height=0.8in]{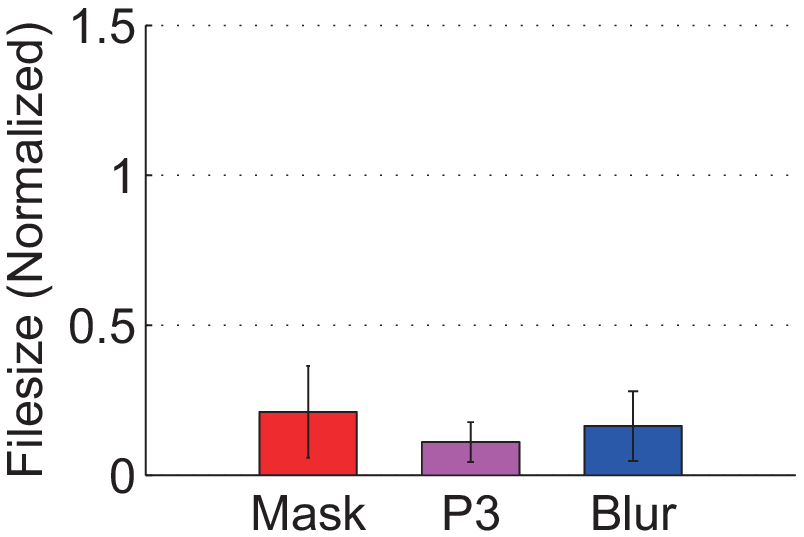}}
   \ \ \ \ \ \ \ \
  \subfigure[Public+Secret]{
  \label{fig:storage:total}
   \includegraphics[width=0.2\linewidth, clip,height=0.8in]{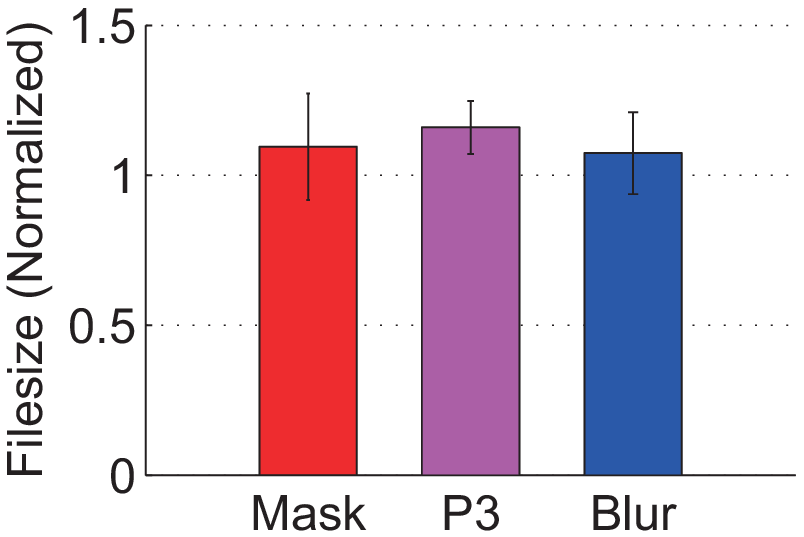}}

    \vspace{-10pt}\caption{Storage cost of three methods for the LFW dataset. (ROP is defined as faces).
    The cost is normalized by the original image file size.}\vspace{-10pt}
  \label{fig:storage_cost}
\end{figure*}

\begin{figure*}[t]
  \centering
   \subfigure[Public Image]{
    \label{fig:storage:pubH}
    \includegraphics[width=0.2\linewidth, clip,height=0.8in]{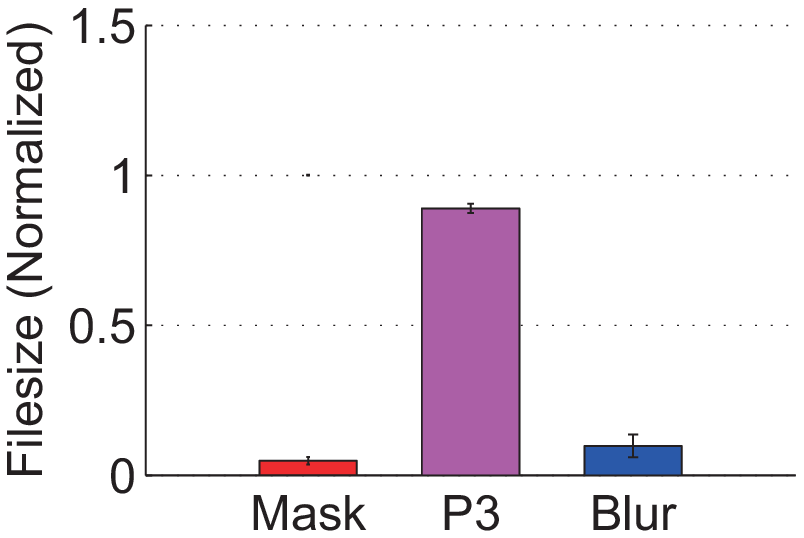}}
    \ \ \ \ \ \ \ \
  \subfigure[Secret Image]{
    \label{fig:storage:priH}
   \includegraphics[width=0.2\linewidth, clip,height=0.8in]{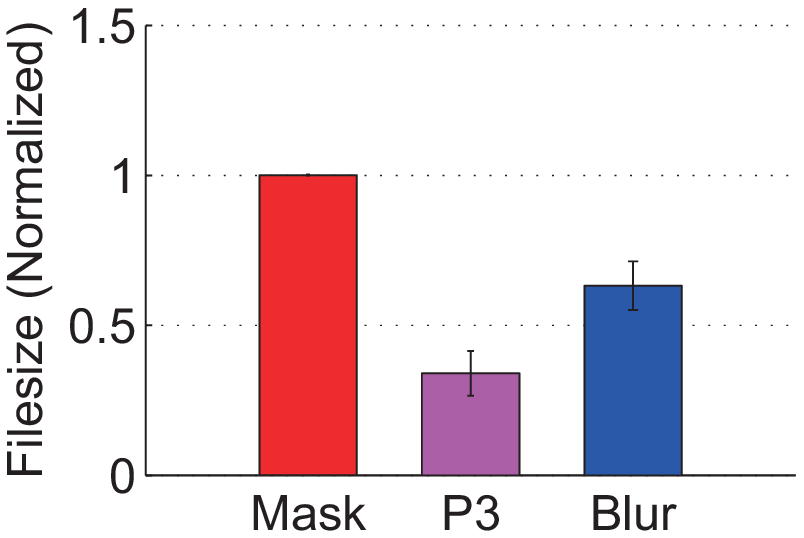}}
   \ \ \ \ \ \ \ \
  \subfigure[Public+Secret]{
  \label{fig:storage:totalH}
   \includegraphics[width=0.2\linewidth, clip,height=0.8in]{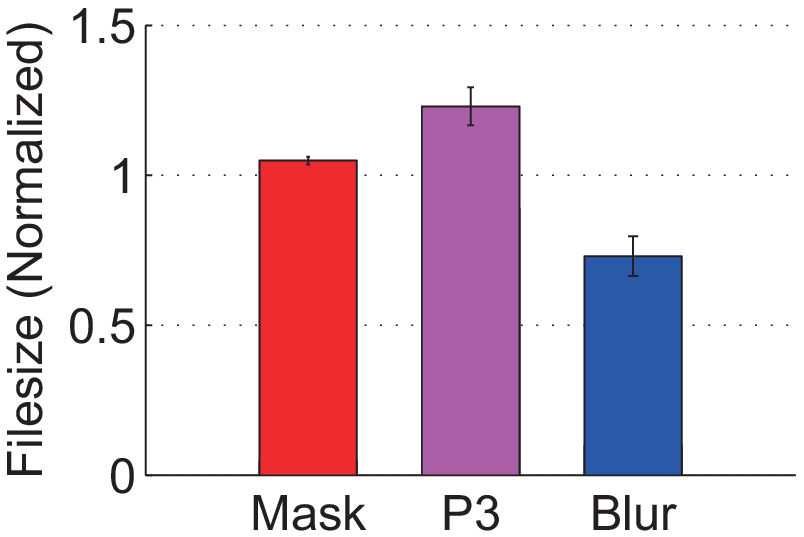}}

  \vspace{-10pt}\caption{Storage cost of three methods for the Holiday dataset. (ROP is defined as the whole image.  The cost is normalized by the original image file size.}\vspace{-18pt}
  \label{fig:storage_costH}
\end{figure*}

We also compare the computation cost
 and storage cost of three methods.
Figure~\ref{fig:concealtime} illustrates the CDF
 of run time for processing each image with three methods.
On average, Mask has the minimum computation cost with $0.0002$s per image in LFW Dataset, and $0.02$s per image in Holiday Dataset;
Blur needs $0.037$s for LFW Dataset and $0.68$s for Holiday Dataset;
P3 needs $0.35$s for LFW Dataset and $24.9$s for Holiday Dataset.
This result also confirms that
 protecting the entire image
is much more expensive than protecting the sub-regions of the image.
Figure~\ref{fig:storage_cost} and \ref{fig:storage_costH}
 present the normalized storage cost of three methods for LFW and Holiday.
The sizes of Blur-processed images are only $73\%$ of the original ones in Holiday dataset on average.
In conclusion,
 Mask and Blur outperforms P3 in computation performance
 while Blur has the best storage performance, therefore \ourprotocol uses Blur as the default method.

\CUTXY{
\paragraph{Privacy of Private Part}
In the private bag,
 the sensitive information include the secret images of ROPs,
 text vectors (\eg, tags),
 feature descriptors and two pairs of keys.
The secret images of ROPs are encrypted with
 the owner's Paillier public key.
The feature descriptors and text vectors are encoded into
  garbled XOR gates, each has only one wire for the input of request feature vector.
The output of the gate is encrypted by the owner's Paillier public key with
 random number.
If the Paillier cryptosystem is semantically secure
 then the secret images, feature descriptors and text vectors are secure against anyone without
 the owner's secret key.

The secret key $SK$ and other keys are all encrypted by the owner defined access structure.
The encryption method CP-ABE is proved secure against
 untrusted storage server and collusion attacks \cite{waters2011ciphertext},
\ie only the requestor with authorized attributes
 is able to decrypt $SK$.
Therefore, \ourprotocol can protect the private bag from
 the \emph{Unauthorized ROP Access} adversary.

\paragraph{Privacy For Requestor}
Besides the privacy of image,
 in \ourprotocolNSP, we also consider the privacy of requestor.
The request feature vector is encoded by garbled $K^0$ and $K^1$.
The order of $K^0$ and $K^1$ is hidden
 by the owner's defined access structure.
The request feature vector will be input into the garbled XOR gates
 to computing the distance between vectors privately.
Since the output of the circuit is secure against unauthorized parties,
 a malicious party cannot tell the input from the output.
As a result,
 the request entry is secure from unauthorized parties.
For the request's retrieved items,
 randomly added indices are used to bind any other parties from the
 $k$-out-of-$n$ selection of the request.
By protecting both the request entry and retrieval items,
 \ourprotocol protects the request and ROP from \emph{ROP Inference}
  adversary.
But a request feature vector could be decode by other authorized parties.

}
\vspace{-0.05in}
\subsection{Search Accuracy}
\vspace{-0.05in}
In \ourprotocolNSP,
 the search procedure follows exactly the same vector-based similarity comparison as typical image search technologies (\eg, \cite{Peker-2011}).
 Also, the accuracy loss introduced by the fixed point representation is almost negligible (less than $\frac{1}{{base}^{scale}}$ in each value where $base$ is often 10 and $scale$ is greater than 5), therefore \ourprotocolNSP~provides a comparable accuracy as existing image search techniques.

\vspace{-0.05in}
\subsection{Client Side Performance}
\vspace{-0.05in}
\begin{table}[t]
\caption{Microbenchmarks}\vspace{-10pt}
\label{table:time}
\centering
\subtable[Image pre-process (LFW/Holiday)\label{table:imagecost}]
{
\centering
\begin{footnotesize}
\begin{tabular}{|l|c|c|c|}
\hline
\multicolumn{4}{|c|}{Laptop (sec)}\\
\hline
& Mean & Min & Max\\
\hline
Detect feature & $0.28$ / $8.7$& $0.17$/$1.56$ & $1.21$/$12.03$\\
\hline
Separate ROP & $0.037$/$0.68$ & $0.009$/$0.07$ & $0.266$/$1.26$\\
\hline
Encrypt $\secret\left(\rop(I)\right)$ &  $0.001$/$0.038$ & $0.001$/$0.018$ & $ 0.008$/$ 0.054$\\
\hline
\multicolumn{4}{|c|}{Smartphone (sec)}\\
\hline
Detect feature  & $0.46$/$15.6$  & $0.21$/$4.32$ & $1.85$/$22.7$\\
\hline
Separate ROP & $0.08$/$1.53$ & $0.024$/$0.19$& $0.37$/$2.73$\\
\hline
Encrypt $\secret\left(\rop(I)\right)$ & $0.005$/$0.057$ & $0.001$/$0.028$& $0.015$/$ 0.13$\\
\hline
\end{tabular}
\end{footnotesize}\vspace{-10pt}
}
\subtable[Image search (average run time)\label{table:searchcost}]{
\centering
\begin{footnotesize}
\begin{tabular}{|l|c|c|}
\hline
\multicolumn{3}{|c|}{Laptop (sec)}\\
\hline
~~~~~~~~~~~~~~~~$\ourprotocolNSP$ & 64 dimension & 128 dimension\\
\hline
Encrypt Vector (owner) & $1.02$ & $2.01$ \\
\hline
Encode Vector (querier) & $0.55$ & $1.12$ \\
\hline
Decrypt Distance & $0.016$ & $0.016$\\
\hline
~~~~~~~~~~~~~~$\ourprotocolNSP_{bin}$ & 64 dimension & 128 dimension\\
\hline
Encrypt Vector (owner) & $0.51$ & $1.03$ \\
\hline
Encode Vector (querier) & $<0.001$ & $<0.001$ \\
\hline
Decrypt Distance & $0.016$ & $0.016$\\
\hline
\multicolumn{3}{|c|}{Smartphone (sec)}\\
\hline
~~~~~~~~~~~~~~~~$\ourprotocolNSP$ & 64 dimension & 128 dimension\\
\hline
Encrypt Vector (owner) & $1.85$ & $3.91$ \\
\hline
Encode Vector (querier) & $0.64$ & $1.37$ \\
\hline
Decrypt Distance &  $0.024$ & $0.024$ \\
\hline
~~~~~~~~~~~~~~$\ourprotocolNSP_{bin}$ & 64 dimension & 128 dimension\\
\hline
Encrypt Vector (owner) & $0.56$ & $1.33$ \\
\hline
Encode Vector (querier) & $<0.001$ & $<0.001$ \\
\hline
Decrypt Distance &  $0.024$ & $0.024$ \\
\hline
\end{tabular}
\end{footnotesize}
}
\vspace{-0.25in}
\end{table}

\subsubsection{Computation Overhead}

For the photo owner,
 the computation overhead  mainly comes from the following operations:
 (1) object detection and descriptor extraction;
 (2) ROP separation by Blur;
 (3) symmetric encryption of secret part;
 (3) descriptor encryption, which are all in the public \& private bag generation.
For the querier,
 the expensive operations include:
 (1) descriptor extraction;
 (2) descriptor encoding;
 (3) distance results decryption;
 (4) similarity calculation, which are all in the cloud-based photo searching.
The time cost for other operations, \eg, fix point presentation conversion, are negligible.
 The cost for search key encryption and decryption by CP-ABE
 can also be ignored,
 since this is a one-time operation for each user which are sub-second.

As microbenchmark tests for each procedure (Table~\ref{table:time}),
Table~\ref{table:imagecost} shows that
 protecting subregions (\eg, faces) of a image
 only takes the owner $0.31$s to extract the descriptor
 and separate the ROP,
 while protecting the whole image takes $9.38$s.
Table~\ref{table:searchcost} presents the computation overhead of main procedures in $\ourprotocolNSP$ and $\ourprotocolNSP_{bin}$.

\paragraph{Public \& Private Bag Generation}
Binary feature vector reduces the owner's computation overhead by half
 to $0.51$s per feature vector.
In a typical scenario in LFW dataset,
 there are only 9 feature vectors for each face.
 if we use 64 dimensional SURF descriptor,
 it takes $9.5$s on laptops and $17$s on smartphones
 to generate the public bag, private bag and search bag.
When we use binary descriptor \cite{Peker-2011},
 the cost is reduced to $4.9$s on laptops and $5.6$s on smartphones, which is a significant reduction.

\paragraph{Cloud-based Photo Searching}
It takes a querier roughly 1s to encode the querying descriptor in $\ourprotocolNSP$.
The run time becomes negligible in $\ourprotocolNSP_{bin}$, and this is especially desirable for mobile devices. After the querier obtains the search result,
 it takes $0.016$s on laptops and $0.024$s on smartphones to decrypt each encrypted distance in both variants.
In the LFW dataset,
 if a querier searches a photo among 1,000 photos,
 it takes $14$s to process the search result on laptops and $22$s
 on the smartphones on average. It is slightly beyond acceptable if owners have hundreds of photos on average.
However, this non-negligible extra overhead comes from the linear search in all photos of an owner with a linear complexity,
and it is promising and not trivial to reduce the complexity with existing optimized search mechanisms such as k-d tree \cite{muja2009fast}.
Thus, the scalability can be achieved using those search algorithms.

\vspace{-0.05in}
\subsubsection{Communication Overhead}
The communication overhead for the image owner
 mainly comes from uploading public bag and private bag to the cloud.
When using Blur to separate ROPs,
as presented in Figure~\ref{fig:storage_cost} and \ref{fig:storage_costH},
the size of the public part is $90\%$ of the original image in LFW Dataset and only $9.8\%$ in Holiday Dataset.
The size of the secret part is $16\%$ in LFW Dataset and $63\%$ in Holiday Dataset.
The average size of encrypted descriptor is 72KB per image for both variants, and this can be further reduced to 690B per image when using a common lossless compression, \eg, ZIP.
As a summary, for the LFW Dataset, the extra communication cost brought by $\ourprotocolNSP$ or $\ourprotocolNSP_{bin}$ is roughly $6\%$ of 
that for system without privacy consideration.
But for the Holiday Dataset, our method actually save the communication cost by $27\%$.

\CUTTaeho{
\ourprotocol can reduce $10\%-90\%$ communication cost
 for uploading the privacy-concealed public image
 to common storage and sharing service, \eg Flicker and Facebook.

The private bags consists of encrypted secret images and descriptors.
For the LFW dataset the secret image is $16\%$
 and for the Holiday dataset, the secret image is $63\%$.
The average size of encrypted real number and binary descriptor
 are both 72 KB for each image.
Using a common lossless compression, \eg ZIP,
 the size of encrypted feature descriptor can be reduced to about 690 Byte for each image.
Overall, compared with system without privacy consideration,
 for the LFW dataset, \ourprotocol costs $6\%$ extra uploading cost for each image,
 and about 690 Byte for encrypted feature descriptor.
For the Holiday dataset,
 \ourprotocol even saves $27\%$ communication cost for uploading images.
}

The communication overhead for uploading the encoded feature descriptors (query)
 is approximately 36 KB in $\ourprotocolNSP$
 and 9 KB in $\ourprotocolNSP_{bin}$, which are reduced to 350B and 90B respectively after compression.
The communication overhead for downloading the similarity result
 is 128B for each compared image in the database, and the one for downloading each image
 is similar to the uploading overhead of the image owner.
Note that,
 to achieve $k$-$n$ oblivious transfer,
 the querier needs to download $(n-k)$ extra images from the search server
 to hide the search pattern, where $n$ can be specified according to the trade-off between privacy and performance.


\vspace{-0.05in}
\subsection{Cloud Side Performance}
\vspace{-0.05in}

On the clouds,
 similarly, the image storage and communication is $6\%$ more for LFW Dataset
 and $27\%$ less for Holiday Dataset.
The main computation overhead is from the
 distance computation.
We evaluated the search performance on laptops,
so the actual performance when deployed in more powerful cloud servers will be significantly improved. Our privacy-preserving distance protocols take nearly $0.18$s to calculate the distance between two real number feature vectors (\ourprotocolNSP)
 and only $0.018$s for binary feature vectors ($\ourprotocolNSP_{bin}$). For well studied objects like faces (9 feature vectors in a descriptor),
 and for each owner, there are usually hundreds of images on the cloud.
The computation time for a laptop to process one request is less than one minute.
When there are large-scale complicated images whose ROPs may contain random objects other than faces,
 the optional optimization methods introduced  may be introduced to reduce the query response time.

\vspace{-0.08in}
\section{Related work}
\vspace{-0.08in}
\label{sec:review}

\paragraph{Image Privacy Protection}
A set of solutions are proposed to mask sensitive contents of images, \eg, human faces,
 to prevent any potential breach of owners' privacy, \eg, \cite{newton2005preserving} and \cite{zhang2005hiding}.
P3 \cite{ra2013p3} proposes to separate an image into a private part and a public part and simply encrypted the private part.
But the produced public parts of those works are of limited utility and disable search on them.
There are some literal works providing privacy-preserving face recognition
 in a face photos database\cite{sadeghi2010efficient}.
Those methods provide privacy protection to the requested images
 as well as the outcome,
 but the result is not secure against photo service provider
 and those works do not consider personal photo storage and sharing.
Supporting privacy-preserving image search with untrusted server is still an open problem.


\paragraph{Privacy Preserving Cloud Services}
Many research efforts have been devoted to provide secure cloud-based 
storage,  sharing and searching services to users.
Those privacy preserving outsourced storage and sharing systems, \eg,\cite{tang2012secure} and \cite{wang2012oruta}, provide
 well access control to private data, but cannot support search on encrypted data.
Searchable encryption is proposed to
 enable secure search over encrypted data via keywords.
But the existing approaches, \eg, \cite{wang2012enabling,li2012toward, wang2012achieving, renprivacy},
 are focus on keywords search by examining the occurrences of the searched terms (or words).
They are not suitable for content-based image search since they cannot measure the distance between encrypted feature vectors.

\paragraph{Privacy-preserving Euclidean Distance}
Euclidean distance can be computed privately among parties
 using secure multi-party computation (SMC) methods \cite{lindell2004proof}.
However,
 it requires online interaction between the image owner
 and queriers, and is unsuitable for the cloud based image service, where the owners are not guaranteed to stay online.
\cite{mukherjee2006privacy} proposes an approach
 using Fourier-related transforms to hide accurate sensitive data and to approximately preserve Euclidean
 distances among them.
It works well for some data mining purposes on common datasets,
 but for feature vectors the distances still reveal information of the objects in images.

\vspace{-0.08in}
\section{Conclusion}
\vspace{-0.08in}
\label{sec:conclusion}
We present a framework \ourprotocolNSP, which enables cloud servers to provide privacy-preserving photo sharing and searching
service to mobile device users who intend to outsource photo management while protecting their privacy in photos.
Our framework not only protects the outsourced photos so that no unauthorized users can access them,
 but also enables users to encode their image search so that the search can also be outsourced to an untrusted cloud server
 obliviously without leakage on the query contents or results.
Our analysis shows the security of the framework,
 and the implementation shows a small storage overhead and communication overhead for both mobile clients and cloud servers.

\vspace{-0.05in}
{
\bibliographystyle{IEEEtran}
\vspace{-0.1in}
\bibliography{ref}
\vspace{-0.1in}
}

\appendix
\section{Security Proof}

Firstly, we prove by the following game that \ourprotocol is secure against adversarial cloud servers who are not authorized for image search.\medskip

\noindent \textsf{Initialize}: System is initialized, and relevant cryptosystems (Paillier's cryptosystem, CP-ABE, OT \etc) are initialized by the challenger $\challenger$. $\challenger$ publishes relevant public keys to the adversary $\adversary$.

\noindent \textsf{Setup}: $\challenger$ generates/encrypts the search keys, and pre-processes/encrypts a set of images $\mathbb{I}$ by the specification of \ourprotocol such that $\adversary$ cannot search on him. Then, he publishes the encrypted search keys and public/private/search bags to $\adversary$.

\noindent \textsf{Phase 1}: $\adversary$ achieves polynomial number of encoded descriptors (encoded with $\challenger$'s search keys) without knowing corresponding original descriptors.

\noindent \textsf{Challenge}: $\adversary$ submits two images $I_0,I_1$ to $\challenger$. $\challenger$ selects a bit $y\in\left\lbrace 0,1 \right\rbrace$ uniformly at random, and generates two sets of encoded feature descriptor $\cipher_1\left(\fpr\left(\fy\right)\right),\cipher_2\left(\fpr\left(\fy\right)\right)$ corresponding to $I_y$ (Section~\ref{sec:cloud_based_search}), which are given to $\adversary$.

\noindent \textsf{Guess}: $\adversary$ gives a guess $y'$ on $y$.\medskip

The advantage of $\adversary$ in this game is defined as
\begin{displaymath}
\textsf{adv}=\text{Pr}\left[
\begin{array}{c}
\adversary\leftarrow \cipher_1\left(\fpr\left(\fy\right)\right),\cipher_2\left(\fpr\left(\fy\right)\right)\\
y'=y
\end{array}\right]-\frac{1}{2}
\end{displaymath}

It is not hard to see this is an adversarial cloud server's advantage, since the game is designed to `mimic' a cloud server's transaction.

\begin{theorem}
Any probabilistic polynomial time adversary (PPTA) has at most negligible advantage in above game.
\end{theorem}

\IEEEproof{
We define two PPTAs $\adversary_1,\adversary_2$, and define their advantages $\textsf{adv}_{i}$ as:
\begin{displaymath}
\textsf{adv}_{i}=\text{Pr}\left[
\begin{array}{c}
\adversary_i\leftarrow\cipher_i\left(\fpr\left(\fy\right)\right)\\
y_i'=y
\end{array}
\right]-\frac{1}{2}
\end{displaymath}

\noindent That is, $\textsf{adv}_i$ is the advantage of $\adversary_i$ when he is only given $\cipher_i\left(\fpr\left(\fy\right)\right)$ and gives a guess $y_i'$ on $y$. Since $\adversary$ is given both adversaries' views, if $\adversary_1,\adversary_2$ agree on the same guess, he will also give the same guess, otherwise his advantage does not change. Then, we have the following probabilities for four cases:
\begin{displaymath}
\begin{split}
\text{Pr}\left[y'=y|y_1'=y\wedge y_2'=y\right]&=1\\
\text{Pr}\left[y'=y|y_1'=y\wedge y_2'\neq y\right]&=\text{Pr}\left[y'=y\right]\\
\text{Pr}\left[y'=y|y_1'\neq y\wedge y_2'=y\right]&=\text{Pr}\left[y'=y\right]\\
\text{Pr}\left[y'=y|y_1'\neq y\wedge y_2'\neq y\right]&=0
\end{split}
\end{displaymath}

Since $\adversary_1$ and $\adversary_2$ gives their guesses based on independent views, we have
\begin{displaymath}
\text{Pr}\left[y_1'=y\wedge y_2'=y\right]=\text{Pr}\left[y_1'=y\right]\cdot \text{Pr}\left[y_2'=y\right]
\end{displaymath}
which also applies to the other three cases. Given those conditional probabilities, the total probability is ($\textsf{adv}=\text{Pr}\left[y'=y\right]-\frac{1}{2}=1-\text{Pr}\left[y'\neq y\right]-\frac{1}{2}$):

\CUTTaeho{
According to the definition of the advantages,
\begin{displaymath}
\textsf{adv}=\text{Pr}\left[y'=y\right]-\frac{1}{2}=1-\text{Pr}\left[y'\neq y\right]-\frac{1}{2}
\end{displaymath}
}

\begin{displaymath}
\begin{split}
\text{Pr}\left[y'=y\right]&=\frac{1}{2}+\textsf{adv}\\
&=1\cdot \left(\frac{1}{2}+\textsf{adv}_1\right)\left(\frac{1}{2}+\textsf{adv}_2\right)\\
&+\left(\frac{1}{2}+\textsf{adv}\right)\left(\frac{1}{2}+\textsf{adv}_1\right)\left(\frac{1}{2}-\textsf{adv}_2\right)\\
&+\left(\frac{1}{2}+\textsf{adv}\right)\left(\frac{1}{2}-\textsf{adv}_1\right)\left(\frac{1}{2}+\textsf{adv}_2\right)\\
&+0
\end{split}
\end{displaymath}

\noindent which leads to
\begin{displaymath}
\textsf{adv}=\frac{\left(\frac{1}{2}+\textsf{adv}_1\right)\left(\frac{1}{2}+\textsf{adv}_2\right)}{\frac{1}{2}-2\textsf{adv}_1\textsf{adv}_2}-\frac{1}{2}
\end{displaymath}

Both Paillier's cryptosystem and CP-ABE are proved to be semantically secure against chosen plaintext attack (SS-CPA)\footnote[3]{CP-ABE is proved to achieve IND-CPA, which implies SS-CPA} \cite{paillier1999public,bethencourt2007ciphertext}. Therefore, $\adversary$ does not have a significant chance to get $SK,\vr$ in $\abe\left(PK,SK,\vr\right)$ or $\fpr\left(\fy\right)$ in $\cipher_2\left(\fpr\left(\fy\right)\right)$, which means $\textsf{adv}_2$ is negligible. Recall that the function family $x\rightarrow \mu x\mod n$ is $\epsilon$-pairwise independent for negligible $\epsilon$, and $\mu^{-1}x\mod n$ is close to uniform in $\mathbb{Z}_n$. Therefore, he does not have a significant chance to get $\fpr\left(\vy_j\left(k\right)\right)$ in $\cipher_1\left(\fpr\left(\vy_j\left(k\right)\right)\right)$ either, which implies a negligible $\textsf{adv}_1$. Since both $\textsf{adv}_1,\textsf{adv}_2$ are negligible, $\textsf{adv}$ is negligible too.
}

Besides adversarial cloud servers, we assumed malicious users in our adversarial model. However, unauthorized malicious users are not as threatening as cloud servers since they never get involved in any transaction with valid users. All they can do is to try man-in-the-middle attacks sniff the search results, which can be trivially prevented with secure communication channel.

\end{document}